\documentclass[aps,prd,reprint,
superscriptaddress, 
nofootinbib,preprintnumbers]{revtex4-2}  
\usepackage{graphicx}  
\usepackage{dcolumn}   
\usepackage{bm}        
\usepackage{amssymb,amsfonts,amsmath,physics}   
\usepackage[dvipsnames]{xcolor}
\usepackage{slashed}
\usepackage{multirow}
\usepackage{dsfont}
\usepackage{booktabs}
\usepackage{comment}
 \usepackage{inputenc}

\usepackage[bookmarks, breaklinks, colorlinks,urlcolor=blue, citecolor=red, 
linkcolor=blue]{hyperref}

\usepackage[normalem]{ulem}

\usepackage{orcidlink}

\def\bar {\overline}

\newcommand{\bmt}{\begin{pmatrix}}
\newcommand{\emt}{\end{pmatrix}}
\newcommand{\ba}{\begin{array}{c}}
\newcommand{\ea}{\end{array}}
\newcommand{\be}{\begin{equation}}
\newcommand{\ee}{\end{equation}}
\newcommand{\bea}{\begin{eqnarray}}
\newcommand{\eea}{\end{eqnarray}}

\newcommand{\bi}{\begin{itemize}}
\newcommand{\ei}{\end{itemize}}

\newcommand{\baz}{\begin{array}{cc}}
\newcommand{\mathsym}[1]{{}}

\newcommand{\bt}{\begin{tabular}}
\newcommand{\et}{\end{tabular}}

\newcommand{\benu}{\begin{enumerate}}
\newcommand{\eenu}{\end{enumerate}}

\newcommand{\Msun}{M_\odot}
\newcommand{\Mstar}{M_\star}

\newcommand{\Rstar}{R_\star}

\newcommand{\MBH}{M_\text{BH}}
\newcommand{\RSch}{R_\text{Sch}}

\newcommand{\pFi}{p_{F,i}}
\newcommand{\pFn}{p_{F,n}}

\newcommand{\kinFn}{\varepsilon_{F,n}}
\newcommand{\kinFi}[1]{\varepsilon_{F,#1}}
\newcommand{\mieff}[1]{m_{#1}^{\rm eff}}

\newcommand{\g}{{\rm \,g}}
\newcommand{\s}{{\rm \,s}}
\newcommand{\cm}{{\rm \,cm}}

\newcommand{\km}{{\rm \,km}}

\newcommand{\K}{{\rm \,K}}

\newcommand{\GeV}{{\rm \,GeV}}
\newcommand{\TeV}{{\rm \,TeV}}
\newcommand{\yr}{{\rm \,yr}}

\newcommand{\Gyr}{{\rm \,Gyr}}

\newcommand{\rth}{r_\text{th}}
\newcommand{\fMB}{f_{\rm MB}}
\newcommand{\fFD}{f_{\rm FD}}
\newcommand{\mbeff}{m_i^{\rm eff}}
\newcommand{\sigmathi}{\sigma^\text{th}_{i\chi}}
\newcommand{\sigmathn}{\sigma^\text{th}_{n\chi}}
\newcommand{\mstar}{m_i^*}

\newcommand{\Tstareq}{T_{\rm eq}}

\begin{document}
\preprint{FERMILAB-PUB-25-0539-T}

\title{From Capture to Collapse: Revisiting Black Hole formation by \\ Fermionic Asymmetric Dark Matter in Neutron Stars}

\author{Sandra Robles\,\orcidlink{0000-0002-6046-8217}}
\email[E-mail: ]{srobles@fnal.gov}
\affiliation{Astrophysics Theory Department, Theory Division, 
Fermi National Accelerator Laboratory, Batavia,
Illinois 60510, USA}
\affiliation{Kavli Institute for Cosmological Physics, University of Chicago, Chicago, Illinois 60637, USA}

\author{Drona Vatsyayan\,\orcidlink{0000-0002-6868-3237}}
\email[E-mail: ]{drona.vatsyayan@ific.uv.es}
\affiliation{Instituto de Física Corpuscular (CSIC-Universitat de València),
Parc Científic UV, C/Catedrático José Beltrán, 2, E-46980 Paterna, Spain}
\affiliation{Departament de Física Teòrica, Universitat de València, 46100 Burjassot, Spain}

\author{Giorgio Busoni\,\orcidlink{0000-0002-8527-0768}}
\email[E-mail: ]{giorgio.busoni@adelaide.edu.au}
\affiliation{ARC Centre of Excellence for Dark Matter Particle Physics, 
Department of Physics, University of Adelaide, South Australia 5005, Australia}


\begin{abstract} 
Fermionic asymmetric dark matter (ADM) can be captured in neutron stars (NSs) via scatterings with the star constituents. The absence of dark matter  annihilation due to its asymmetric nature leads to ADM accumulation in the NS core, potentially reaching densities sufficient to exceed the Chandrasekhar limit and trigger its gravitational collapse into a black hole (BH), eventually consuming the NS from within. Therefore, the existence and observation of old neutron stars  provide a means to constrain the properties of ADM. We revisit previous constraints on the   mass and scattering cross section off neutrons of fermionic ADM across a class of models. We critically examine common simplifying approximations used in the literature to derive these limits. 
Our analysis includes improved treatments of dark matter capture, thermalization, BH formation, accretion, and evaporation.  We find that previous results can be relaxed by a few orders of magnitude once these effects are properly accounted for.
\end{abstract}

\maketitle



\section{Introduction}
\label{sec:intro}

Neutron stars (NSs) as cosmic laboratories are increasingly becoming popular probes to constrain the properties of dark matter (DM), offering complementarity with direct detection experiments. 
DM accelerated to quasi-relativistic speed by the strong gravitational potential of NSs, can accumulate in the NS after being captured by scattering with the NS constituents, depositing its kinetic energy onto the NS. This can have several consequences. 
For instance, if DM self-annihilates, it can also transfer its mass energy to the NS core~\cite{Kouvaris:2007ay,Bertone:2007ae,Kouvaris:2010vv}.

A popular alternative to the weakly interacting massive particles (WIMP) scenario is that of the asymmetric dark matter (ADM), originally motivated by the cosmic coincidence, i.e., the relation between baryonic and DM densities: $\rho_{\rm DM} \sim 5 \rho_{\rm B}$. ADM consists only of the remnant asymmetric population of DM (similar to the SM baryons) determined by an initial asymmetry in the dark sector, hence, the DM annihilations of the standard WIMP scenario are negligible (see Refs.~\cite{Kaplan:2009ag,Petraki:2013wwa,Zurek:2013wia} for a review). If enough ADM is accumulated in the NS core, it can trigger gravitational collapse and lead to the formation of a black hole (BH) that can eventually destroy the NS. Therefore,  observations of old NSs could be used to constrain the interaction strength and mass of DM~\cite{Goldman:1989nd,Kouvaris:2010jy,McDermott:2011jp,Kouvaris:2011fi,Guver:2012ba,Kouvaris:2012dz,Bell:2013xk,Bramante:2013nma,Bramante:2014zca,Bramante:2015dfa,Garani:2018kkd,Dasgupta:2020dik,Giffin:2021kgb,Lu:2024kiz,Liu:2024qbe,Dutta:2024vzw,Basumatary:2024uwo}.

While such earlier studies have made use of certain assumptions or approximations, here, we show how relaxing those assumptions, and carefully including the relevant effects/corrections concerning the physics of NSs can alter the previous results. The approach of such studies usually involves four steps: (i) Capture: DM scattering with NS material and becoming gravitationally bound to the star; (ii) thermalization: the captured DM undergoes further collision with NS matter until it reaches thermal equilibrium close to the star's center;  (iii) self-gravitation: enough  thermalized DM accumulates in the NS core to become  self-gravitating; and
 (iv) NS collapse:  the  self-gravitating  DM would collapse  when its gravitational energy exceeds its kinetic energy. This corresponds to the Chandrasekhar limit in the case of fermionic DM. Once this limit is reached, a BH forms in the NS core, potentially leading to the NS destruction.  
 
In each of these steps, we discuss some important caveats and modifications to the standard approach, in particular, improved capture rates to include relativistic effects, the effect of Pauli blocking and multiple scatterings~\cite{Bell:2020jou,Bell:2020lmm,Bell:2020obw,Anzuini:2021lnv}. The effect of Pauli blocking on the thermalization timescale for both light and super-heavy DM is discussed, leading to a different behavior in the latter case compared to previous results. Further, we also discuss the effects of NS temperature evolution for thermalization timescales. We find that, while they are not significant for super heavy DM, they can modify the thermalization time for DM masses less than $10^7$ GeV. We review the conditions for BH formation,  discuss the relativistic and quantum effects~\cite{Giffin:2021kgb} for BH accretion, and give caveats for standard Bondi accretion in this context. The limits on the critical BH mass below which it evaporates are discussed by carefully considering the evaporation rate, depending on the BH temperature. We also comment on the consequences of suppressed BH evaporation. Our results show that previous results can be relaxed by a few orders of magnitude. 

In this work, we are interested in the case of fermionic ADM candidates. We consider an effective field theory (EFT) approach to characterize the interactions of a Dirac DM with the standard model (SM) quarks in order to obtain the constraints that can be applied to a wide range of fermionic DM models. The outline of the work is as follows: In Section~\ref{sec:NSs}, we discuss the structure, composition and cooling of NSs;  Section~\ref{sec:capture} deals with the improved capture rates for DM in NSs.  The thermalization timescale and the effect of NS cooling is discussed in Section~\ref{sec:therm}. The destruction of NSs including the approach for BH accretion and evaporation is discussed in Section~\ref{sec:dest}. Finally, we present our results in Section~\ref{sec:results} and summarize in Section~\ref{sec:conclusions}.

\section{Neutron Stars}
\label{sec:NSs}

The core of a NS harbors not only degenerate neutrons, but also protons, electrons, and muons in beta equilibrium. Its exact composition remains unknown, though. 
At the extreme densities found in the inner regions of the core of massive neutron stars,  hyperons, kaon condensates, or even quark matter may exist. 

To parametrize this uncertainty in the NS composition as well as the microphysics relevant for this study, we model the NS interior by making a  sole assumption, an equation of state. 
As in Refs.~\cite{Bell:2020obw,Anzuini:2021lnv,Bell:2023ysh}, we use the quark-meson coupling (QMC) model~\cite{Guichon:2018uew,Motta:2019tjc} to obtain radial profiles for the target number densities, Fermi energies $\kinFi{i}$, effective masses $\mbeff$ for every NS constituent $i$, and the speed of sound $c_s$, as well as $B(r)$ (the coefficient of the time part of the Schwarzschild metric), by solving the Tolman-Oppenheimer-Volkoff (TOV) equations~\cite{Tolman:1939jz,Oppenheimer:1939ne}.\footnote{Note that by solving the TOV equations we are implicitly assuming a spherical, symmetric and a slow rotating NS.} For an example of the radial profiles used in this work, see Ref.~\cite{Anzuini:2021lnv}; values at the center and surface for  key properties are listed in Table~\ref{tab:NSconfig}. 

The specific QMC model used here allows for the presence of hyperons in the inner NS core. They appear when the NS mass surpasses $\sim1.7\Msun$. 
The first hyperon to appear is $\Lambda^0$, when the neutron chemical potential equals that of $\Lambda^0$. Next, $\Xi^-$ appears when the neutron plus electron chemical potentials equal that of the $\Xi^-$. $\Xi^0$ is also present in the heaviest configuration $\Mstar\geq1.9\Msun$, though in less significative quantities. Thus, hyperons lower matter pressure by replacing high momentum neutrons with low momentum hyperons.

\begin{figure}[tb]
    \centering
    \includegraphics[width=\columnwidth]{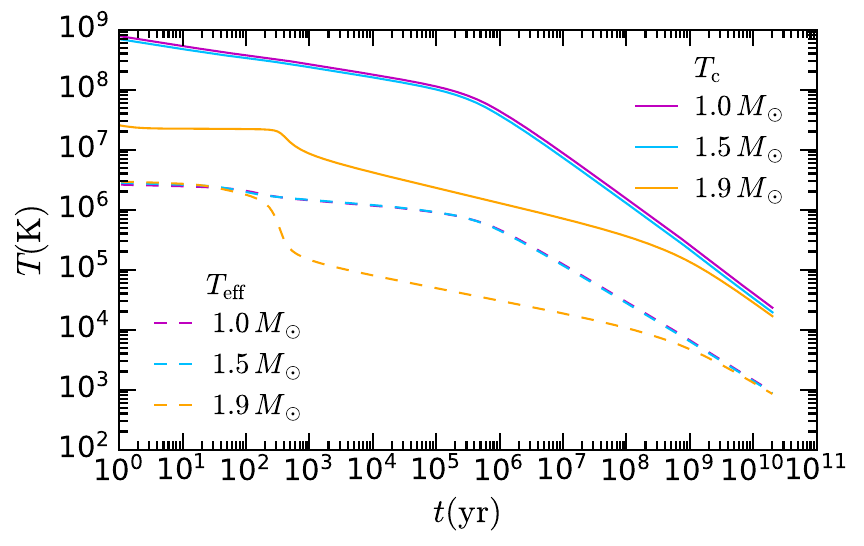}
    \caption{The cooling of an isolated non-rotating neutron star: Solid and dashed curves core ($T_c$) and effective ($T_{\rm eff}$) temperature evolution with time of the three benchmark NS masses that we work with.}
    \label{fig:cool}
\end{figure}

The NS core accounts for $\sim99\%$ of the star mass, while the thin crust plays a major role observationally. The outermost layers of the crust act as a heat blanketing envelope, thereby determining the effective surface temperature, $T_\text{eff}$, inferred by observations, which as can be seen in Fig.~\ref{fig:cool} is orders of magnitude lower than that of the core,  $T_\text{c}$.

\begin{table*}[tb]
\centering
\begin{tabular}{|l|c|c|c|c|c|c|c|c|c|c|}
\hline
\bf EoS & $\rho_c=\rho(0)$ & $P_c=P(0)$ & $\Mstar$  & $\Rstar$ &  $B(0)$ & $B(\Rstar)$ & $c_s(0)$ & $\mieff{n}(0)$ & $\pFn(0)$ & $\kinFn(0)$  \\
& [$\rm g/cm^3$] & [$\rm dyne/cm^2$] & $[\Msun]$ & [km] & & & [c]  &  [GeV] & [GeV] & [GeV]\\
\hline
\bf QMC-1 & $5.692\times 10^{14}$& $4.343\times10^{34}$ & 1.000 & 13.044 & 0.580 & 0.772 & 0.480  &  0.591 & 0.407 &  0.127\\
\bf QMC-2 & $8.108\times 10^{14}$& $1.070\times10^{35}$ & 1.500 & 12.847 & 0.408 & 0.653 & 0.591 &  0.540 & 0.448 & 0.162 \\
\bf QMC-3 & $1.803\times10^{15}$& $3.246 \times10^{35}$ &1.900 & 12.109 & 0.235 & 0.535 & 0.412 &  0.463 & 0.446 & 0.180   \\
\hline
\end{tabular} 
\caption{Properties of the benchmark NS configurations as determined in Refs.~\cite{Bell:2020obw,Anzuini:2021lnv}, using the QMC equation of state and varying the central baryon number density.  From left to right, these quantities are central density, central pressure, NS mass $\Mstar$; radius $\Rstar$; $B(0)$ and $B(\Rstar)$ related to the escape velocity at the center and surface of the NS, respectively;   the speed of sound, neutron effective mass $\mieff{i}(0)$, Fermi momentum, and Fermi energy $\kinFi{i}(0)$  at the NS center.}
\label{tab:NSconfig}
\end{table*}

The cooling of a non-rotating, non-magnetic neutron star that does not accrete standard matter is modeled using 
the \texttt{NSCool} code, neglecting superfluidity \cite{Ootes:2016oib}. Note that in the absence of a cooling model for the later stages of a NS, \texttt{NSCool} relies on an extrapolation of the relation between the core temperature and the surface temperature  found in Ref.~\cite{Nomoto:1987}. The core $T_\text{c}$ and effective temperature $T_\text{eff}$ for the three NS benchmark configuration masses (see Table~\ref{tab:NSconfig}) are shown in Fig.~\ref{fig:cool} where we have fixed the age of the star at ${\cal O}(10^{10})$ years. Photon cooling becomes dominant around $10^5\yr$ for the QMC-1 ($\Mstar=1\Msun$) and QMC-2  ($\Mstar=1.5\Msun$) benchmarks, respectively; see position of the knee around these timescales in both $T_c$ and $T_\text{eff}$. Note that the QMC-3 model ($\Mstar=1.9\Msun$) contains hyperons that make the NS cool faster during its early stages. In this case, direct Urca processes are active, and the end of the neutrino cooling era is delayed with respect to the lighter NS configurations.

On the other hand, observations of Gyr old pulsars in the ultraviolet and optical bands suggest that an ideal scenario as that in Fig.~\ref{fig:cool} does not hold and that NSs can be reheated.  
Mechanisms that can potentially reheat  old isolated NSs,  without requiring any new physics are, e.g., accretion from the interstellar medium, weak deviations from beta equilibrium, \textit{rotochemical heating}~\cite{Reisenegger:1994be,Reisenegger:1997,Fernandez:2005cg,Gonzalez:2010}, frictional motion of superfluid neutron vortices, \textit{vortex creep}~\cite{Alpar:1984,Shibazaki:1989,Larson:1998it}, 
 rotation-induced deep crustal heating (similar to rotochemical heating, but operating in the crust) \cite{Gusakov:2015kaa}, and dissipation of a strong magnetic field \cite{Haensel:1990,Miralles:1998na,Pons:2008fd}. The first two rotation powered mechanisms can satisfactorily explain the high temperatures observed in old millisecond pulsars (MSPs)~\cite{Gonzalez:2010}, and they are also consistent with the upper bound on the surface temperature of the coldest observed NS J2144–3933~\cite{Guillot:2019ugf} $T_\text{eff}<42000\K$, a $\sim 0.3\Gyr$ old, slowly rotating, isolated pulsar.

\section{Capture}
\label{sec:capture}
\noindent The total amount of DM captured in the neutron star is simply given by
\begin{equation}
    N_\chi (t) = C t\,,
\end{equation}
where C is the capture rate. In the optically thin limit it is given by~\cite{Bell:2020jou}
\begin{align}
    C_\text{thin} =\frac{4\pi \rho_\chi}{m_\chi}&\int_0^\infty d\mu_\chi \frac{f_{\rm MB}(\mu_\chi)} {\mu_\chi} \nonumber \\ &\times\int_0^{R_\star} dr\, r^2 \frac{\sqrt{1-B(r)}}{B(r)}\Omega^{-}(r), \label{eq:capturethin}
\end{align}
where $r$ is the radial distance, $\rho_\chi \sim 0.4 {\rm~GeV}\,{\rm cm}^{-3}$ is the local DM density, $m_\chi$ is the DM mass, $\mu_\chi$ is the DM velocity far away from the NS, and $f_{\rm MB}$ is the assumed Maxwell-Boltzmann distribution of relative velocities between the NS and DM far away from it. $B(r)$ is the radial profile related to the escape velocity $v_{\rm esc}^2 (r) = 1 -B(r)$. $\Omega^-(r)$ is the interaction rate with fixed DM kinetic energy [$K_\chi = m_\chi(1/\sqrt{B(r)}-1)$ relevant for a DM particle falling into the NS], and is given by \cite{Bell:2020obw,Anzuini:2021lnv}
\begin{align}
    &\Omega^{-}(r)=\frac{1}{32\pi^3}\int dt\,d E_i\,ds \frac{s|\bar{M}(s,t,m_i^{\rm eff})|^2}{s^2 -[(m_i^{\rm eff})^2-m_\chi^2]^2}\frac{E_i}{m_\chi}\nonumber\\
    & \times\sqrt{\frac{B(r)}{1-B(r)}}
     \frac{f_{\rm FD}(E_i,r)(1-f_{\rm FD}(E'_i,r))}{\sqrt{[s-(m_i^{\rm eff})^2]^2 - 4(m_i^{\rm eff})^2 m_\chi^2}}\,,
\end{align}
where $s$ and $t$ are Mandelstam variables, $\mbeff(r)$ is the effective mass of the baryonic target $i$, $E_i$ and $E_i'$ are the initial and final energy of the target, respectively;  $\fFD$ is the Fermi Dirac distribution, see Ref.~\cite{Anzuini:2021lnv} for further details. The exact expression for the squared matrix element  $|\overline{M}_{i\chi}(s,t,\mbeff)|^2$  for this interaction can be found in Table~\ref{tab:operators}. 

The DM accretion rate in Eq.~\ref{eq:capturethin} assumes that DM capture can occur within the stellar interior. However, if the DM-nucleon cross section is larger than a threshold value $\sigmathi$, the DM will be captured closer to the surface, entering the optically thick regime, and eventually will saturate to the so-called geometric limit. This is when all DM particles traversing the star are captured, regardless of the value of the cross section. To account for this effect, we introduce in Eq.~\ref{eq:capturethin} an optical factor $\eta$,  which is a function of the DM optical depth $\tau_\chi$ and depends on the trajectory followed by the DM until it scatters with a NS constituent~\cite{Bell:2020jou}. For single scattering capture, it reads 
\begin{equation}
\eta(r) = \frac{1}{2 J_\text{max}^2} \int_0^{J_\text{max}} \frac{J \, dJ}{\sqrt{J_\text{max}^2-J^2}} \left[e^{-\tau_\chi^-(r,J)}+e^{-\tau_\chi^+(r,J)}\right], 
\label{eq:optfactor}
\end{equation}
where $J$ is the DM angular momentum and $J_\text{max}$ is the maximum possible angular momentum at a distance $r$ from the center of the NS; this is
 \begin{equation}
    J_\text{max}^2(r) = \frac{1-B(r)}{B(r)} m_\chi^2 r^2.
\end{equation} 
Note that in Eq.~\ref{eq:optfactor} we are averaging over the two equally probable trajectories a DM particle can follow to reach a shell of radius $r$ within the NS. 

If the DM mass is larger than a certain threshold value, which we denote $\mstar$, more than one collision is required for the DM to become gravitationally bound to the NS. The exact value of this threshold mass depends on the specific target particle and on the radial coordinate. The values $\mstar(r)$ were calculated in Ref.~\cite{Anzuini:2021lnv}. 
Moreover, if the cross section is also large enough (larger than $\sigmathi$),  the star opacity cannot be neglected. In this case, the opacity factor that corrects Eq.~\ref{eq:capturethin} is
\begin{align}
\eta^\text{multi}(r) =& \frac{1}{2J_\text{max}^2n^*(r)} \int_0^{J_\text{max}} \frac{J \, dJ}{\sqrt{J_\text{max}^2-J^2}} \nonumber\\
&\times\left(e^{-\frac{\tau_\chi^-(r,J)}{n^*(r)}}+e^{-\frac{\tau_\chi^+(r,J)}{n^*(r)}}\right),
\label{eq:optfactorlargem}
\end{align}
where  $n^*(r)$ can be interpreted as the number of scatterings required to capture a DM particle. Thus, for $n^*(r)\rightarrow1$, we recover Eq.~\ref{eq:optfactor}. 
The precise value of  $n^*(r)$ was found to be \cite{Bell:2020jou}
\begin{equation}
n^*(r) = \frac{1}{1-\exp\left[-m_i^*(r)/m_\chi\right]}.  
\end{equation}
In the limit of very large DM mass, it can be approximated with 
\begin{equation}
n^*(r) \sim \frac{m_\chi}{m_i^*(r)}.
\end{equation}

Then, to account for capture via multiple scattering, we introduce  the new optical factor $\eta^\text{multi}(r)$, Eq.~\ref{eq:optfactorlargem}  in Eq.~\ref{eq:capturethin}, 
\begin{align}
C_\text{opt} =  &\frac{4\pi\rho_\chi}{m_\chi} 
\int_0^\infty du_\chi \frac{\fMB(u_\chi)}{u_\chi} \nonumber\\
& \times \int_0^{\Rstar} dr\, r^2 \frac{\sqrt{1-B(r)}}{B(r)}  \eta^\text{multi}(r) \Omega^{-}(r).
\label{eq:capturemult} 
\end{align}

If all the DM passing through the NS is captured, one can obtain an absolute upper bound on the amount of DM captured using the geometric capture rate~\cite{Bell:2018pkk} 
\begin{equation}
    C_{\rm geom} = \frac{\pi R_\star^2[1-B(R_\star)]}{v_\star B(R_\star)}\frac{\rho_\chi}{m_\chi}{\rm Erf}\left(\sqrt{\frac{3}{2}}\frac{v_\star}{v_d}\right)\,,
    \label{eq:capturegeom}
\end{equation}
where $v_\star$ is the NS velocity, $v_d$ is the DM velocity dispersion, and 
\begin{equation}
    B(R_\star) = 1 - \frac{2GM_\star}{c^2 R_\star}\,.
\end{equation}
This is the optically thick regime, where the NS can be regarded as a hard sphere.

\begin{figure}
    \centering
    \includegraphics[width=\columnwidth]{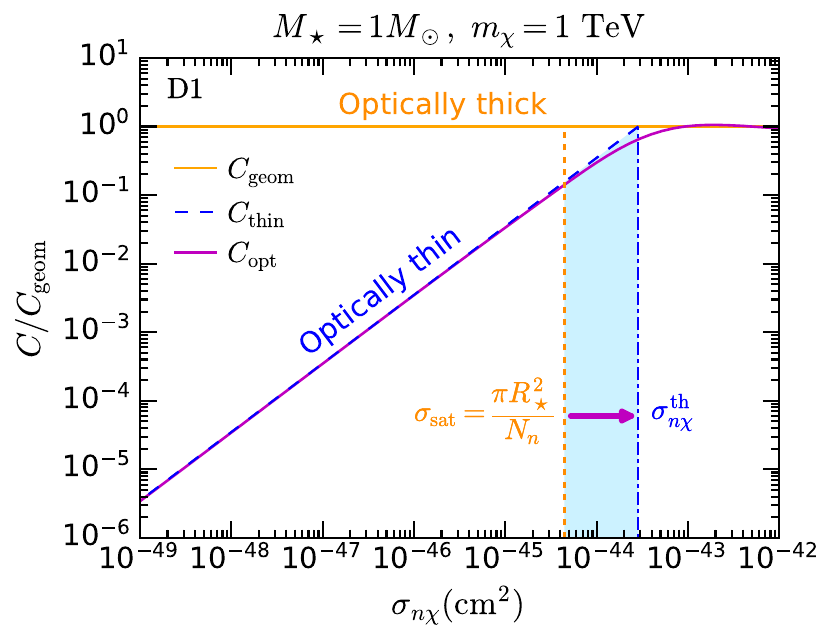}    
    \includegraphics[width=\columnwidth]{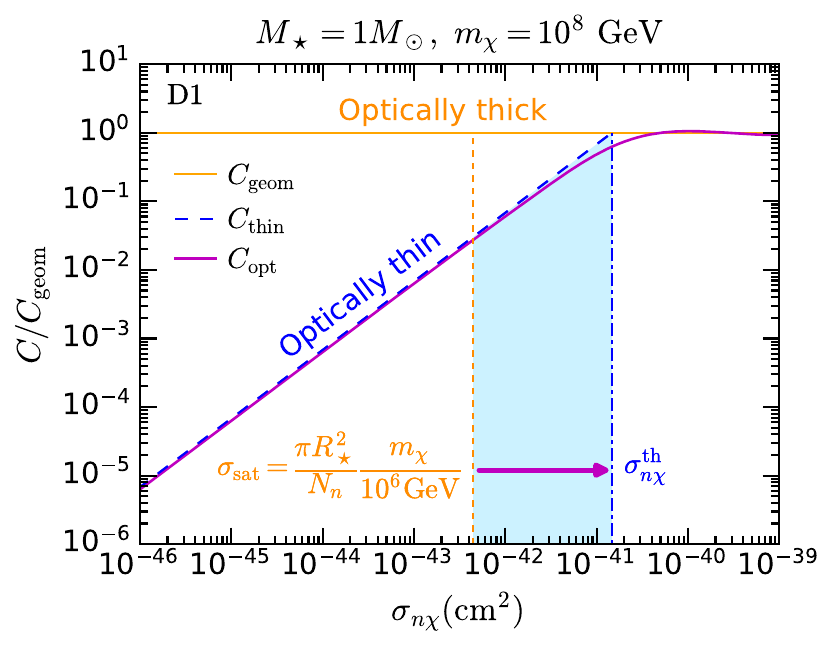} 
    \caption{Capture rate as a function of the DM-neutron cross section for a $1\Msun$ NS and DM of mass $m_\chi =1\TeV$ (top) and $m_\chi =10^8\GeV$ (bottom)  with scalar-scalar interactions with quarks (D1), assuming the equation of state QMC. The vertical dot-dashed blue line denotes the threshold cross section for DM-neutron scattering, which marks out the transition from the optically thin to the optically thick regime. For comparison, we show the  saturation cross sections commonly used in the literature. }
    \label{fig:Csigma}
\end{figure}

\begin{table*}[!htb]
    \centering
    \begin{tabular}{|c|c|c|c|c|c|}
        \hline
        Name & Operator & $g_q$ & $g_i^2(t)$ & $|\overline{M}(s, t, m_i)|^2$ & Dominant term $t_{\text{term}}$ \\
        \hline
        D1 & $\bar{\chi} \chi \; \bar{q} q$ & $\frac{y_q}{\Lambda_q^2}$ & $\frac{c_i^S(t)}{\Lambda_q^4}$ & $g_i^2(t) \frac{(4m_\chi^2 - t)(4m_\chi^2 - \mu^2 t)}{\mu^2}$ & $t^0$ \\
        D2 & $\bar{\chi} \gamma^5 \chi \; q q$ & $i \frac{ y_q}{\Lambda_q^2}$ & $\frac{c_i^S(t)}{\Lambda_q^4}$ & $g_\chi^2(t) \frac{t (\mu^2 t - 4m_\chi^2)}{\mu^2}$ & $t^1$ \\
        D3 & $\bar{\chi} \chi \; \bar{q} \gamma^5 q$ & $i \frac{ y_q}{\Lambda_q^2}$ & $\frac{c_i^P(t)}{\Lambda_q^4}$ & $g_\chi^2(t) t (t - 4m_\chi^2)$ & $t^1$ \\
        D4 & $\bar{\chi} \gamma^5 \chi \; \bar{q} \gamma^5 q$ & $ \frac{ y_q}{\Lambda_q^2}$ & $\frac{c_i^P(t)}{\Lambda_q^4}$ & $g_\chi^2(t) t^2$ & $t^2$ \\
        D5 & $\bar{\chi} \gamma_\mu \chi \; \bar{q} \gamma^\mu q$ & $\frac{1}{\Lambda_q^2}$ & $\frac{c_i^V(t)}{\Lambda_q^4}$ & $2g_i^2(t)\frac{2(\mu^2+1)^2 m_\chi^4 -4(\mu^2+1)\mu^2 s m_\chi^2+\mu^4(2s^2+2st+t^2)}{\mu^4}$ & $t^0$ \\
        D6 & $\bar{\chi} \gamma_\mu \gamma^5 \chi \; \bar{q} \gamma^\mu q$ & $\frac{1}{\Lambda_q^2}$ & $\frac{c_i^V(t)}{\Lambda_q^4}$ & $2g_i^2(t)\frac{2(\mu^2-1)^2 m_\chi^4 -4\mu^2 m_\chi^2 (\mu^2 s + s+t) +\mu^4(2s^2+2st+t^2)}{\mu^4}$ & $t^0$ \\
        D7 & $\bar{\chi} \gamma_\mu \chi \; \bar{q} \gamma^\mu \gamma^5 q$ & $\frac{1}{\Lambda_q^2}$ & $\frac{c_i^A(t)}{\Lambda_q^4}$ & $2g_i^2(t)\frac{2(\mu^2-1)^2 m_\chi^4 -4\mu^2 m_\chi^2 (\mu^2 s + s+t) +\mu^4(2s^2+2st+t^2)}{\mu^4}$ & $t^0$ \\
        D8 & $\bar{\chi} \gamma_\mu \gamma^5 \chi \; \bar{q} \gamma^\mu \gamma^5 q$ & $\frac{1}{\Lambda_q^2}$ & $\frac{c_i^A(t)}{\Lambda_q^4}$ & $2g_i^2(t)\frac{2(\mu^4+10\mu^2+1) m_\chi^4 -4(\mu^2+1)\mu^2 m_\chi^2(s+t)+\mu^4(2s^2+st+t^2)}{\mu^4}$ & $t^0$ \\
        D9 & $\bar{\chi} \sigma_{\mu\nu} \chi \; \bar{q} \sigma^{\mu\nu} \gamma^5 q$ & $\frac{1}{\Lambda_q^2}$ & $\frac{c_i^T(t)}{\Lambda_q^4}$ & $8g_i^2(t)\frac{4(\mu^4+4\mu^2+1) m_\chi^4 -2(\mu^2+1)\mu^2 m_\chi^2(4s+t)+\mu^4(2s+t)^2}{\mu^4}$ & $t^0$ \\
        D10 & $\bar{\chi} \sigma_{\mu\nu} \gamma^5 \chi \; \bar{q} \sigma^{\mu\nu} q$ & $\frac{1}{\Lambda_q^2}$ & $\frac{c_i^T(t)}{\Lambda_q^4}$ & $8g_i^2(t)\frac{4(\mu^2-1)^2 m_\chi^4 -2(\mu^2+1)\mu^2 m_\chi^2(4s+t)+\mu^4(2s+t)^2}{\mu^4}$ & $t^0$ \\
        \hline
    \end{tabular}
    \caption{Dimension 6 EFT operators for Dirac DM-quark couplings and their squared matrix elements, which depend on two parameters $m_\chi$ and the energy scale $\Lambda_q$. $y_q$ are the quark Yukawa couplings, $\mu=m_\chi/\mbeff$, and the hadronic coefficients $c_i(t)$ can be found in Ref.~\cite{Anzuini:2021lnv}. The last column corresponds to the dominant term in the matrix relevant for the calculation of the thermalization time~\cite{Bell:2023ysh}.}
    \label{tab:operators}
\end{table*}

In Fig.~\ref{fig:Csigma}, we showcase the transition from the optically thin  Eq.~\ref{eq:capturethin} to the optically thick regime Eq.~\ref{eq:capturegeom}, as well as the estimation of the DM-neutron threshold cross section for a scalar mediator (operator D1, see Table~\ref{tab:operators}). This is the interaction usually assumed to derive bounds from direct detection experiments. 
It is worth remarking that not even for the most simple interactions, scalar mediated, the DM-neutron scattering cross section can be taken as a constant when computing capture rates in a NS, since  the hadronic matrix element in $|\bar{M}(s,t,m_i^{\rm eff)})|$ contains momentum-dependent terms that are not negligible and the radial dependence of the mass of the targets make this assumption unrealistic. We can compute a cross section in the non-relativistic limit as a reference value that can be compared to direct detection experiments; for D1 this cross section reads
\begin{equation}
    \sigma_{n\chi} = \frac{c_N^Sm_\chi^2m_n^2}{\pi\Lambda_q^4(m_\chi+m_n)^2},
    \label{eq:sigmanonrelD1}
\end{equation}
where $c_N^S$ is the squared matrix element for the scalar operator (for the exact value, see Refs.~\cite{Bell:2018pkk,Anzuini:2021lnv}).

To estimate the DM-neutron threshold cross section $\sigmathn$ (dot-dashed blue lines), we simply equate Eq.~\ref{eq:capturethin}, $C(\Lambda_q)$, to the expression for the capture rate in the geometric limit Eq.~\ref{eq:capturegeom}, derive $\Lambda_q$. and then using Eq.~\ref{eq:sigmanonrelD1} convert the cutoff scale $\Lambda_q$ to a non-relativistic cross section. As we can see in Figure~\ref{fig:Csigma}, the optically thin limit is a very good approximation for the capture rate, compare magenta with dashed  blue lines. Furthermore, we compare our definition of $\sigmathn$ with the naive estimation, dubbed saturation cross section, $\sigma_\text{sat}=\pi\Rstar^2/N_n$ (dashed orange lines), with $N_n$ the number of neutron targets. The latter being most commonly used for stellar objects in the single scattering regime. As we can immediately notice,  this saturation cross section underestimates the onset of the optically thick regime by about an order of magnitude for the particular case of a $1\Msun$ NS.

In the case for which multiple collisions are necessary to capture local ambient DM, previous estimations found  the threshold DM mass for this to occur to be  ${\cal O}(10^6\GeV)$~\cite{Bramante:2017xlb} or $m_T/v^2_\text{halo}$~\cite{Joglekar:2020liw} (where $m_T$ is the target mass and $v_\text{halo}\simeq220\km \s^{-1}$ is the DM velocity far from the star).  When properly accounting for the physics relevant to the extreme regime found in  neutron stars, in particular momentum-dependent hadronic matrix elements and baryon effective masses, this threshold mass is lowered down to  ${\cal O}(10^5\GeV)$~\cite{Anzuini:2021lnv}. 
This implies  that the saturation cross section 
$\sigma_\text{sat}=\pi\Rstar^2/N_n \times m_\chi/10^6\GeV$~\cite{Baryakhtar:2017dbj} is not a good approximation for the DM-neutron geometric cross section, as it can immediately be seen in Fig.~\ref{fig:Csigma}, bottom panel.

In light of these results, in what follows we shall use Eq.~\ref{eq:capturethin} to calculate the capture rate, taking care that the DM-neutron cross section cannot exceed the threshold value.

After being captured, the DM continues to interact with the NS constituents. In particular, very light DM is susceptible to up-scatter in these further collisions, i.e., to gain enough energy in a scattering to escape the NS. A threshold DM mass below which this is relevant can be estimated using the prescription outlined in Ref.~\cite{Bell:2020lmm}.  This is the so-called evaporation mass and is highly dependent on the star temperature, being roughly ${\cal O}(100T_\text{c})$~\cite{Bell:2020lmm} for an ${\cal O}(10^{10})$ years old NS. The precise value depends on the NS configuration and the specific target particle. Finite temperature effects on the calculation of the capture rate are also expected to be important below the evaporation mass.
Thus, evaporation can be safely neglected in the region of the parameter space we are analyzing.

\section{Thermalization}
\label{sec:therm}

\begin{figure}
    \centering
    \includegraphics[width=\columnwidth]{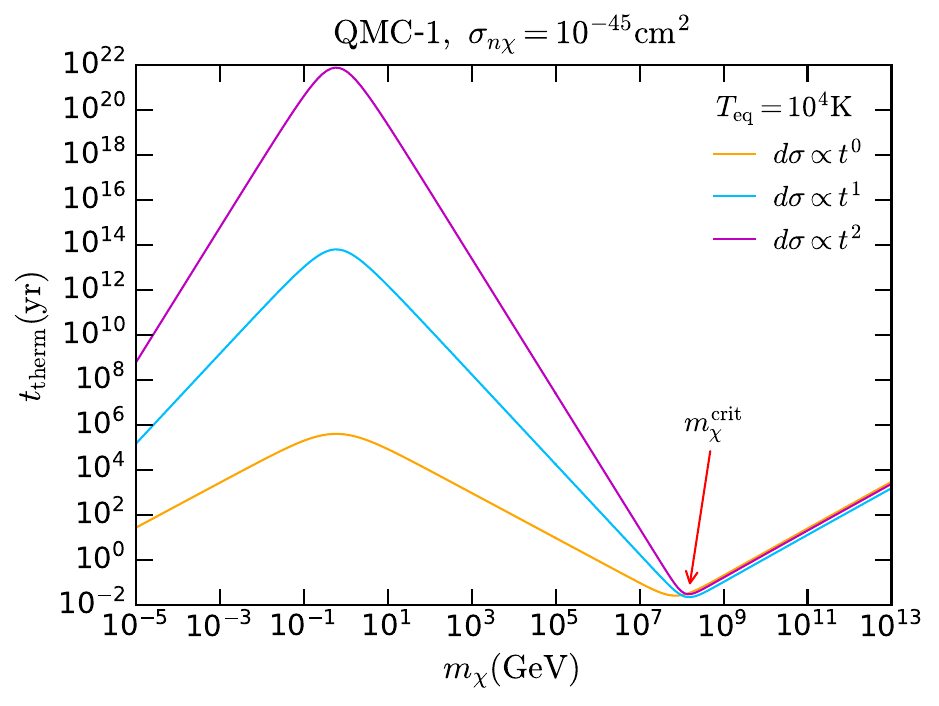}
    \caption{Thermalization time as a function of the DM mass for DM-neutron cross sections such that $d\sigma_{n\chi}\propto t^n$, $n=0,1,2$. The cross section on the surface is set to $\sigma_{n\chi}=10^{-45}\cm^2$, and the equilibrium temperature at the center of a $1\Msun$ NS with EoS QMC to $\Tstareq=10^4\K$. }
    \label{fig:ttherm}
\end{figure}

Captured DM orbits the NS and continues to scatter with the NS material, loosing energy in each collision until it finally  settles down in the center of the NS, reaching thermal equilibrium with the NS constituents at a temperature $\Tstareq$.

Depending on the DM mass, Pauli blocking can affect the  thermalization process; the critical mass for this to occur is~\cite{Bell:2023ysh}
\begin{equation}
     m_\chi \lesssim\frac{2\kinFi{i}(2 \mieff{i} + \kinFi{i})}{\Tstareq} =m_\chi^{\rm crit}. 
    \label{eq:mucrit}
\end{equation}
For DM masses smaller than $m_\chi^{\rm crit}$, Pauli blocking delays the thermalization process. 
Thus, the thermalization timescale for this DM mass range is determined by the last few scatterings occurring in the vicinity of the NS center, which take much longer than a second, depending on the DM mass. 
In particular, the suppression on the interaction rate due to Pauli blocking is larger 
 for momentum-dependent cross sections. 
Hence, to compute the thermalization  timescale in this mass range, it is sufficient to retain the term with the lowest power of momentum transfer $t$ in the cross section~\cite{Bell:2023ysh}; see Table~\ref{tab:operators}, last column. 
For the case in which the DM-target cross section can be taken as a constant, $d\sigma_{i\chi}\propto t^0$, the thermalization time is given by~\cite{Bell:2023ysh}
\begin{align}
    t_{\rm therm}^{(n=0)}\sim \frac{147\pi^2 m_\chi}{16(m_i^{\rm eff}(0)+m_\chi)^2\sigma_{i\chi}^{n=0}\Tstareq^2},
    \label{eq:tthermn0}
\end{align}
where the baryon effective mass is taken at the NS center. 
The approximate expressions for the thermalization time for $d\sigma_{i\chi}\propto t^n$, $n=1,2$, as well how to obtain approximate expression of this timescale for the dimension 6 EFT operators can be found in Ref.~\cite{Bell:2023ysh}. 

For $m_\chi\gtrsim m_\chi^{\rm crit}$, this is when the DM is heavy enough that has enough kinetic energy during most of the thermalization process so that Pauli blocking never affects thermalization, the collisions that lead to reaching this equilibrium state are spaced by a fraction of a second. In this regime, the thermalization timescale for $d\sigma_{i\chi}\propto t^0$ is well approximated by\footnote{Here we are correcting the expression given in Ref.~\cite{Bell:2023ysh}.}
\begin{equation}
     t_{\rm therm}^{(n=0)}  \sim \frac{3 \pi^2 m_\chi}{8 (\mbeff(0))^2 \kinFi{i}^2 \sigma_{i\chi}^{n=0} }\log\left[\frac{m_\chi}{T_{\rm eq}}\left(\frac{1}{\sqrt{B(\Rstar)}}-1\right)\right]. 
\label{eq:tthemheavy0text}
\end{equation}

In the transition region between the above-mentioned regimes, Pauli blocked (discrete process) to non Pauli blocked (continuous process) thermalization, the thermalization time can be estimated as the sum of both expressions, Eqs.~\ref{eq:tthermn0} and~\ref{eq:tthemheavy0text} for the case of constant DM-neutron cross section in the following way. First, the DM loses energy in a quasi-continuous process  until it reaches the critical kinetic energy
\begin{equation}
    K_\chi^{\rm crit} \simeq\frac{2\kinFi{i}(2 \mieff{i} + \kinFi{i})}{m_\chi} . 
    \label{eq:Tcrit}
\end{equation}
The time to reach this state is given by Eq.~\ref{eq:tthemheavy0text}, replacing $\Tstareq$ with $K_\chi^\text{crit}$. 
After crossing this barrier, the remaining collisions are Pauli suppressed with longer and longer time intervals between scatterings until the DM is finally in equilibrium with the NS material at a temperature $\Tstareq$. The time for the second part of the process is given by Eq.~\ref{eq:tthermn0}, after subtracting the time required to reach $K_\chi^\text{crit}$ under the same approximation (this negative term is obtained by replacing $\Tstareq$ with $K_\chi^\text{crit}$ in Eq.~\ref{eq:tthermn0}). 
Explicitly,  for $d\sigma_{i\chi}\propto t^0$ this reads
\begin{widetext}
\begin{equation}
 t_{\rm therm}^{(n=0)}  \sim \frac{3 \pi^2 m_\chi}{8 (\mbeff(0))^2 \kinFi{i}^2 \sigma_{i\chi}^{n=0} }\log\left[\frac{m_\chi}{ K_\chi^{\rm crit}}\left(\frac{1}{\sqrt{B(\Rstar)}}-1\right)\right]
 + \frac{147\pi^2 m_\chi}{16(m_i^{\rm eff}(0)+m_\chi)^2\sigma_{i\chi}^{n=0}} \left[ \frac{1}{\Tstareq^2} - \frac{1}{(K_\chi^{\rm crit})^2}\right]. 
\end{equation}
\end{widetext}

In Fig.~\ref{fig:ttherm}, we can observe the three regions, Pauli blocked, transition and non-Pauli blocked thermalization, for a fixed value of the DM-neutron scattering cross section on the NS surface. This is done for different interactions, such that $d\sigma_{n\chi}\propto t^n$, $n=0,1,2$. For an equilibrium temperature of $\Tstareq=10^4\K$, the transition between the aforementioned regimes occurs at around $m_\chi^{\rm crit}\sim4\times10^8 \GeV$ for our NS benchmark QMC-1. This is marked for a change of slope, from that DM mass on, the thermalization time will monotonically increase with the DM mass. 
These approximations are in good agreement with the numerical results presented in Ref.~\cite{Bell:2023ysh}. 

\begin{figure}
    \centering
    \includegraphics[width=\columnwidth]{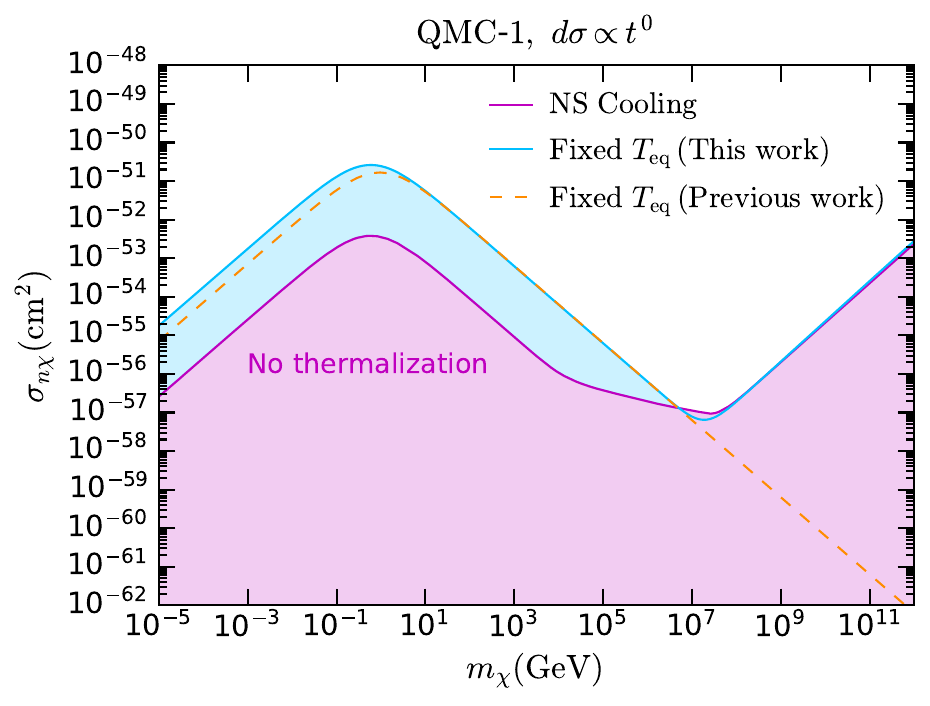}
    \includegraphics[width=\columnwidth]{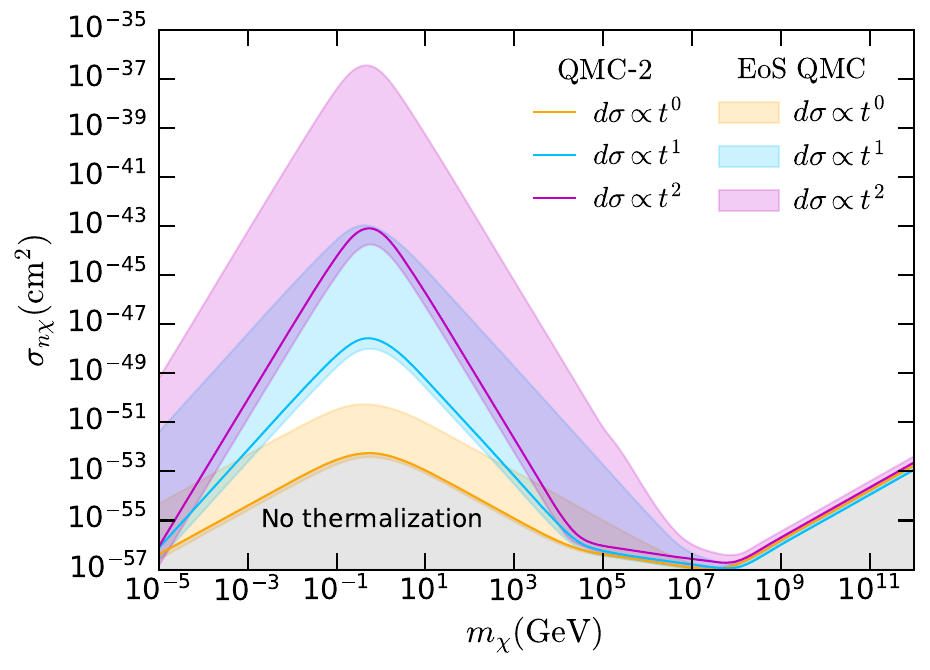}    
    \caption{Top: Region in the $m_\chi - \sigma_{n\chi}$ plane in which DM does not thermalize (magenta shaded region) for $d\sigma_{n\chi}\propto t^0$, i.e., $t_{\rm therm}^{(n=0)}>t_\star$. In the white region, the captured DM fully thermalizes, whereas in the light blue shaded region, a fraction of total captured DM thermalizes. The orange dashed curve represents the thermalization results from Ref.~\cite{Garani:2018kkd}. Bottom: Cross section for no thermalization for our different NS benchmarks for $d\sigma_{n\chi}\propto t^n$, $n=0,1,2$.}
    \label{fig:notherm}
\end{figure}

If the thermalization timescale for a given DM mass and DM-target cross section is larger than the age of the NS, the DM never thermalizes. Therefore, one can determine the region in the $m_\chi - \sigma_{i\chi}$ plane where the captured DM particles do not thermalize within the NS core; thus no meaningful constraints can be derived on them from  NS observations. In previous studies, a fixed temperature $T_{\rm eq}$ was used to derive these constraints~\cite{McDermott:2011jp,Garani:2018kkd}, and the effects of NS cooling were not taken into account. 

As these thermalization timescales depend on the temperature $T_{\rm eq}$, we can equate it with the core temperature of the NS for the thermalized DM, which changes with time due to the NS cooling (see the discussion above). Therefore, at earlier times, when the temperature of the NS core was much higher than the present temperature, the thermalization timescale is shorter as $t_{\rm therm} \propto T_{\rm eq}^{-1}$.   Hence, it is possible that, for a given value of $m_\chi$ and $\sigma_{i\chi}$, we have $t_{\rm therm} < t_\star$ at earlier times, however, at later times $t_{\rm therm} > t_\star$. In this case, a small fraction of the total captured DM can still thermalize.

The region where $t_{\rm therm}^{(n=0)} > t_\star$ at all times, including the NS cooling effects, is shown in magenta in the top panel of Fig.~\ref{fig:notherm} for the case of neutron targets. The DM thermalization is not very efficient in this shaded region and we refer to this as the ``no thermalization'' region.\footnote{It was shown in Ref.~\cite{Garani:2018kkd} that a small fraction of captured DM can thermalize even for $t_{\rm therm}>t_\star$, once the DM energy distribution function is properly taken into account. However, this small thermalized fraction has no observable consequences.} We also show the maximum DM-neutron scattering cross section for which the DM thermalizes assuming a fixed final temperature (light blue line)  ${\cal O}(10^4~{\rm K})$ (the exact value is taken from Fig.~\ref{fig:cool} at $t=10^{10}\yr$) for comparison, as well as the result using the estimate for thermalization time given in Ref.~\cite{Garani:2018kkd}. 

It can be seen that the shape of the light blue curve is similar to the shape of the curve in Fig.~\ref{fig:ttherm}. Since the thermalization time increases for very heavy DM mass, namely for $m_\chi\gtrsim 4\times10^8\GeV$, in order to have $t_{\rm therm}<t_\star$, we are pushed towards larger cross sections, in contrast with the results of Ref.~\cite{Garani:2018kkd}. Moreover, we see that the light blue and magenta curves coincide for $m_\chi \gtrsim 10^8~{\rm GeV}$ due to the logarithmic dependence of the thermalization time on the equilibrium temperature for heavy DM masses (see Eq.~\ref{eq:tthemheavy0text}). Thus, the inclusion of cooling effects is more prominent for $m_\chi \leq m_{\rm crit}$, where $t_{\rm therm} \propto T_{\rm eq}^{-(n+2)}$. The plateau region in between the two regimes is due to the dependence of $m_\chi^{\rm crit}$ on $T_{\rm eq}$ and the total thermalization time being given by the sum of Eqs.~\ref{eq:tthermn0} and \ref{eq:tthemheavy0text}. 

In the white region above the fixed $T_{\rm eq}$ curve, the thermalization process is highly efficient, and all captured DM is thermalized, whereas the light blue shaded region corresponds to the partial thermalization of captured DM. As we shall see in the section below, the relevant DM-neutron scattering cross sections corresponding to fermionic DM that can lead to BH formation, lie in the region where the captured DM is fully thermalized. 

In the bottom panel of Fig.~\ref{fig:notherm}, we illustrate the dependence of the DM-neutron cross section for no thermalization, for $d\sigma_{n\chi}\propto t^n$, $n=0,1,2$.  
The width of each band denotes this, from QMC-1 (smallest $\sigma_{n\chi}$) to QMC-3  (largest $\sigma_{n\chi}$); QMC-2 is shown as a solid line. The dramatic difference between the results for QMC-2 and QMC-3 is due to the presence of hyperons, which drastically reduces the NS cooling times as depicted in Fig.~\ref{fig:cool}. This results in larger  thermalization timescales than in a NS of the same mass that does not harbor hyperons in its core. Given a fixed period of time, such as $t_\star$ in Fig.~\ref{fig:notherm}, longer thermalization times imply larger scattering cross sections for no thermalization with respect to stars with no hyperon content. 
Note that, for the next section, ADM particles collapsing into a black hole, DM should have first achieved thermal equilibrium with the NS material. This is, e.g., for the QMC-1 NS the DM-neutron scattering cross section is required to be larger than the values depicted by the solid magenta line in the top panel. 

It is worth remarking here that we have not considered thermalization by scattering with superfluid nucleons. As shown in Ref.~\cite{Bertoni:2013bsa}, if neutrons in the NS core are in a superfluid phase, this will prevent thermalization from being achieved due to the highly limited  kinematics of the neutron superfluid. However, it would still be possible for the DM to reach thermal equilibrium via scatterings with leptons, which proceeds as described in this section. 

\section{Destruction of neutron stars}
\label{sec:dest}

As time elapses, a population of asymmetric DM will grow in the innermost part of the NS core. After  thermal equilibrium has been reached, we can assume that the DM forms a sphere, very close to the star's center. Then, the total potential that includes both the NS gravitational potential and the contribution of the DM cloud  in the absence of self-interactions between DM particles  reads~\cite{Steigerwald:2022pjo}
\begin{align}
 U_{N_\chi} = &-4\pi m_\chi  \int_0^{\infty} \phi(r) \, n_\chi(r,t) r^2 dr \nonumber  \\
 &- 4\pi G m_\chi \int_0^{\infty} M_\chi(r,t) n_\chi(r,t) r dr ,
 \label{eq:UN}
\end{align}
where $\phi(r)$ is the NS gravitational potential that close to the NS center can be approximated by $\phi(r)\simeq 2\pi G r^2(\rho_c+3P_c)/3$, with $\rho_c$ and $P_c$ the density and pressure at the NS center, respectively. 
$n_\chi$ is the number density of thermalized DM inside the NS. Provided that the average distance between DM particles $(n_\chi)^{-1/3}$ is larger than their thermal de Broglie wavelength, 
\begin{align}
    \lambda_\text{th} &= \sqrt{\frac{2\pi}{m_\chi \Tstareq}} \nonumber\\
    &= 5.33\times10^{-11}\cm \left[\left(\frac{10^4 \K}{\Tstareq}\right)\left(\frac{1 {\TeV}}{m_\chi}\right) \right]^{1/2},
\end{align}
the DM number density follows a Maxwell-Boltzmann distribution, i.e., 
\begin{align}
   n_\chi(r,t) &= \dfrac{N_\chi(t)\exp\left[{-r^2/r_\chi^2(t)}\right]} {\int_0^{\infty} 4\pi r^2 \exp\left[{-r^2/r_\chi^2(t)}\right]} \nonumber \\ &\simeq \frac{N_\chi(t)}{\pi^{3/2} r_\chi^3(t)} \exp[-r^2/r_\chi^2(t)],   
   \label{eq:nDMiso}
\end{align}
where $r_\chi(t)$ is the radius where the DM particles are confined that decreases over time. 

$M_\chi(r,t)$ is the  DM mass enclosed in a radius $r$ at a time $t$, this is  
\begin{equation}
   M_\chi(r,t)=4\pi m_\chi \int_0^r n_\chi(r',t) r'^2 dr'.  
\end{equation}

Then, the mean potential energy per DM particle is given by~\cite{Bell:2024qmj}
\begin{equation}
U_\chi =  - \pi G (\rho_c+3P_c) m_\chi r_\chi^2- \frac{G m_\chi^2 N_\chi(t)}{\sqrt{2\pi} r_\chi}. 
\label{eq:UDM}
\end{equation}
Using the virial theorem, we find the condition for DM to achieve self-gravitation, 
\begin{equation}
    N_\chi(t) \ge \frac{2\sqrt{2}\pi^{3/2}\rth^3(\rho_c+3P_c)}{3\sqrt{3} m_\chi} = N_\mathrm{self},  \label{eq:Ncrit}
\end{equation}
which is given in terms of the thermal radius $\rth$ usually defined in the literature when the gravitational potential due to the accumulated DM cloud is negligible (second term in Eq.~\ref{eq:UDM}),  
\begin{align}
    \rth &= \sqrt{\frac{3\Tstareq}{\pi G(\rho_c+3P_c) m_\chi}}\\
    &= 3.33{\cm} \left[\left(\frac{\Tstareq}{10^4 {\K}}\right)\left(\frac{1 {\TeV}}{m_\chi}\right) \left( \frac{10^{15}\g/\cm^3}{\rho_c+3P_c/c^2} \right)\right]^{1/2}. 
\end{align}
Note that when self-gravitation is achieved the DM is actually confined to a smaller radius $\rth\rightarrow\rth/\sqrt{3}$. 

Once the condition in Eq.~\ref{eq:Ncrit} is reached, the DM becomes more and more confined, with increasing Fermi momentum, and DM self-gravity dominates the potential energy in Eq.~\ref{eq:UDM}. 
For fermionic DM,  the self-gravitating DM cloud will undergo gravitational collapse as soon as its Fermi energy cannot counterbalance the gravitational energy per particle, where the Fermi energy of the confined DM is
\begin{equation}
    E_{F,\chi} = \sqrt{m_\chi^2+p_{F,\chi}^2}-m_\chi, \qquad
    p_{F,\chi} = (3\pi^2n_\chi)^{1/3}. 
\end{equation}
This leads to the following condition for the onset of gravitational collapse: 
\begin{equation}
    \frac{G m_\chi^2 N_\chi(t)}{\sqrt{2\pi} r_\chi} > E_{F,\chi},
\end{equation}
with $r_\chi<\rth$. 
The Fermi energy satisfies $E_{F,\chi} \leq p_{F,\chi}$. Then, the condition for black hole formation at the center of the NS, 
\begin{equation}\label{eq:nch}
N_\chi(t)> (2)^{3/4} \pi \sqrt{3} \left(\frac{M_\text{Pl}}{m_\chi}\right)^3 = N_\text{Ch},   
\end{equation}
where we have used a Maxwell-Boltzmann DM distribution similar to Eq.~\ref{eq:nDMiso} to calculate the DM Fermi momentum. This accounts for the constant factor that does not appear in previous calculations, where DM particles were assumed to be radially uniformly distributed, and that can amount up to an order of magnitude difference with previous work.   
The radius of the DM sphere $r_\chi$ at the onset of collapse  results in 
\begin{equation}
    r_\chi \leq (2)^{1/4} \sqrt{3\pi} \frac{M_\text{Pl}}{m_\chi^2}=0.88\cm \left(\frac{1 {\TeV}}{m_\chi}\right)^2.
\end{equation}

\subsection{Black hole accretion}

Once a black hole (BH) is formed at the center of the NS, it will accrete NS matter as well as DM. The accretion of standard matter onto a black hole has previously been studied in the context of gas clouds from the interstellar medium being accreted by  stars and compact objects. Note that accreted particles are considered point-like, not self-bound and at rest far away from the star/compact object. It is worth emphasizing here that most of the literature on ADM in NSs employed the spherically symmetric Bondi accretion expression~\cite{Bondi:1952}
\begin{equation}
    \frac{dM}{dt} = \frac{4\pi \lambda_s \rho_\infty G^2 M^2}{c_{s,\infty}^3},
    \label{eq:Bondi}
\end{equation}
where $M$ is the mass of the accreting object,  $\lambda_s$ is a constant that depends on the gas polytropic index $\Gamma$, $P=K\rho^\Gamma$ with $K$ a constant and $P$ the gas pressure. Note that $\lambda_s$ is evaluated at the point where the velocity of the infalling particles matches their speed of sound, i.e., the sonic point. $\rho_\infty$ and $c_{s,\infty}$ are the gas density and speed of sound far away from the object. There exists a relativistic derivation of this equation~\cite{Michel:1972oeq,Shapiro:1983du}, which at lowest order agrees with the Bondi equation, provided that the gas is non-relativistic at infinity, i.e., $c_{s,\infty}\ll1$. As we can see in Table~\ref{tab:NSconfig}, this is clearly not our case.

The case where a point-like mass moves through a gas cloud that behaves as a perfect fluid has also been considered. This is the so-called Bondi-Hoyle accretion~\cite{Bondi:1944,Bondi:1952}
\begin{equation}
    \frac{dM}{dt} = \frac{4\pi \bar{\lambda} \rho_\infty G^2 M^2}{(c_{s,\infty}^2+v^2_\infty)^{3/2}},
    \label{eq:BondiHoyle}
\end{equation}
where $v_\infty$ is the velocity of the accreting object and $\bar{\lambda}$ is an ${\cal O}(1)$ constant, not to be confused with $\lambda_s$. This is actually an interpolated result, first suggested by Bondi~\cite{Bondi:1952} and later corrected by Ref.~\cite{Shima:1985}. However, this is again a Newtonian expression. Relativistic effects are important when the accreting object is a NS or a BH. This case has also been studied, mostly numerically, see Ref.~\cite{Edgar:2004mk} for early studies.

\subsubsection{Relativistic effects}

Here, we are dealing with a small BH at rest at the center of a NS, which feeds on the NS matter, mainly neutrons in the core. Thus, it will remove neutrons from the Fermi surface, which are not at rest. This particular problem has been previously studied \cite{East:2019dxt,Richards:2021upu}, including  general relativistic simulations of the BH, but considering Bondi-like accretion with the NS matter at rest. 
Note that the Fermi velocity of the neutrons is always greater than the speed of sound and, of course, they are gravitationally bound. Therefore, we cannot apply any of the previous treatments for accretion of standard matter to our particular case. We can still assume spherical symmetry and the NS matter as a perfect fluid, assumptions behind the TOV equations, and write down the conservation of mass flux equation
\begin{equation}
    \dfrac{d\MBH}{dt} = \sqrt{B(r)} 4 \pi \mbeff(r) n_i(r) v(r) r^2, 
\end{equation}
where $v$ is the velocity of the baryons, $r$ is the radial coordinate, and we have  included the gravitational redshift factor $\sqrt{B(r)}$.   Accounting for the degeneracy of the NS matter states leads to 
\begin{equation}
    \dfrac{d\MBH}{dt} = \sqrt{B(r)} 4 \pi \mbeff(r)  r^2 \int \frac{p}{E_i} f_p(p) dp, 
\end{equation}
where $f_p$ is the momentum distribution
\begin{equation}
 f_p(p,r) dp = \frac{p^2}{\pi^2}\fFD(E_i,r) dp,
\end{equation}
which in the zero temperature approximation leads to the following accretion rate at a given radius from the NS center: 
\begin{align}
    \dfrac{d\MBH}{dt}\bigg|_A = &\frac{4}{\pi}\sqrt{B(r)} \mbeff(r)  r^2\nonumber\\
    & \times \int_0^{\pFi} \frac{p^3}{\sqrt{p^2+(\mbeff)^2}}  dp. 
    \label{eq:baraccrate}
\end{align} 
We can evaluate this expression at the Schwarzschild radius of the BH $\RSch=2G\MBH/c^2$, and given the smallness of the latter  we can approximate the density and Fermi momentum with their values at the NS center. Eq.~\ref{eq:baraccrate} is similar to the geometric accretion rate given in Ref.~\cite{Autzen:2014tza}. 

\begin{figure}
    \centering
    \includegraphics[width=\columnwidth]{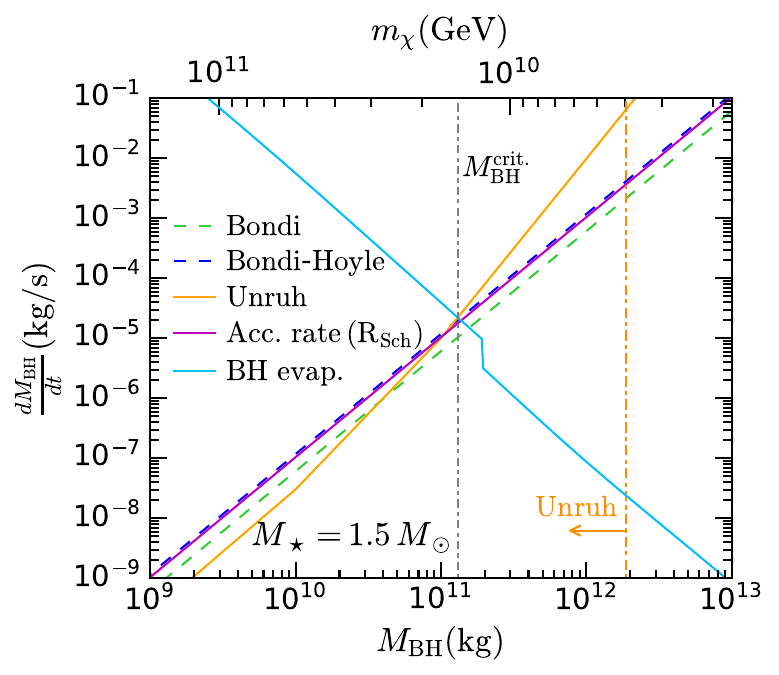}
    \caption{Accretion rates of neutrons in a $1.5\Msun$ NS with EoS QMC. The dot-dashed orange line denotes the BH mass for which $\RSch(\MBH)=1/\pFn$. }
    \label{fig:accrates}
\end{figure}

\subsubsection{Quantum effects}

 In the scenario we are analyzing, for super heavy DM ($m_\chi \gtrsim 10^{9} {\rm~GeV})$, quantum effects arise when the Schwarzschild radius of the formed BH  becomes smaller than the de Broglie wavelength of the particles being accreted, i.e., $R_{\rm Sch} < \lambda_{dB} = 2\pi/\pFi$. This case was originally studied by Unruh in Ref.~\cite{Unruh:1976fm}, where the absorption cross section was found to be 
\begin{align}
   \sigma_U(&\MBH,m_i,v) = \frac{(4\pi^2 G^3\MBH^3m_i)(1+v^2)}{v^2\sqrt{1-v^2}} \nonumber \\
   &\times \left[ 1-\exp\left(- \frac{2\pi G\MBH m_i(1+v^2)}{v^2\sqrt{1+v^2}}\right)\right]^{-1},
   \label{eq:absUnruh}
\end{align}
where $m_i$ and  $v$ are the mass and velocity of the fermion, respectively. This calculation was derived  assuming that the wavelength of the particle being absorbed is larger than the Schwarzschild radius and that particles can be described by free plane waves far away from the BH. 
The latter is not precisely the case of the NS constituents; however, lacking a better approach to tackle this problem, we will take this formalism as a first approximation to deal with potential quantum effects.    
Using Eq.~\ref{eq:absUnruh}, we can compute the accretion rate   in the following way~\cite{Giffin:2021kgb}:  
\begin{equation}\label{eq:bhunruh}
    \dfrac{d\MBH}{dt} \bigg|_U = \frac{\sqrt{B}\mbeff}{\pi^2}  \int_0^{p_F} dp \,  \frac{p^3}{E_i} \sigma_U(\MBH,\mbeff,p),
\end{equation}
where we have dropped the dependence on the radial coordinate for simplicity.

In Fig.~\ref{fig:accrates}, we compare the different accretion rates discussed here. We have computed them considering only accretion of neutrons for the NS benchmark QMC-2 and including the corresponding gravitational redshift factors in Eqs.~\ref{eq:Bondi} and \ref{eq:BondiHoyle}. The Bondi-Hoyle accretion rate (dashed  blue line) has been evaluated for $\bar{\lambda}=1$ and averaged over the velocity, accounting for Pauli blocking. Despite being an empirical approach, it approximates well the accretion rate at the BH Schwarzschild radius, Eq.~\ref{eq:baraccrate}. Conversely, Bondi accretion (dashed green), which is not the right approach for this particular scenario, underestimates it by a factor of $\sim2$. 
The yellow dot-dashed line corresponding to $\MBH^\text{U} \sim 1.9 \times 10^{12}$ kg, is the BH mass below which the Unruh accretion rate  always holds;  recall that this accretion rate assumes $\RSch(\MBH)<1/\pFi$. For heavier BHs the accretion rate is described by Eq.~\ref{eq:baraccrate}. Note that the initial mass of the BH can be converted to the DM mass necessary to form it using Eq.~\ref{eq:nch}, as shown in the secondary x-axis for reference.

\subsubsection{Effects of NS rotation}

Here, we comment on the effect of the rotation of neutron stars on the accretion rate. As discussed in Ref.~\cite{Kouvaris:2013kra}, the rotation of neutron stars and thus that of the infalling matter implies that the conditions of spherically symmetric accretion are not always met (as in the Bondi accretion regime). This is due to the fact that the infalling matter forms a disk around the BH due to its angular momentum, thus suppressing the standard accretion rate.

However, these effects can be countered by including the viscosity effects of the nuclear matter; therefore, the BH accretion rate is suppressed only for $M_B < \MBH < M_c$, where
\begin{align}
    M_c &\simeq 3.9 \times 10^{19} P^{-3}{~\rm kg}\,,\nonumber\\
    M_B &\simeq 3.5 \times 10^{27} \left(\frac{P}{T_5^4}\right)^{1/3}{~\rm kg}\,,
\end{align}
where $P$ is the star rotation period given in seconds and $T_5 = (T_c/10^5 K)$, with $T_c$ as the NS temperature. 

It can be seen that these limits correspond to BH masses much heavier than the initial BH mass formed due to accumulation of DM, and thus correspond to masses toward the final phase of NS destruction. Even for millisecond pulsars, e.g. with $P \sim 5$ ms and $T_c = 10^5$ K, we see that $M_B > M_{c}$; hence, there is no significant suppression and one can proceed with the accretion term given by Eq.~\ref{eq:baraccrate}. 

This suppression, however, can become important for warmer NSs ($T_c \gg 10^5$ K), leading to $M_B  < M_c$. While we track the temperature evolution of a NS in this analysis, the formation of a BH takes place at $t_\star = \mathcal{O}(1-10{\Gyr})$ (as discussed below), at which the NS core has rapidly cooled down to core temperatures ${\cal O}(10^4\K)$, and we are safe from these effects.

\subsection{Black hole  evaporation}

After the initial BH is formed, it continues to feed on the NS matter (and DM) and the rate of change of its mass is governed by either Eq.~\ref{eq:baraccrate} or Eq.~\ref{eq:bhunruh} depending on the regime we are in. If the de Broglie wavelength of the particles being absorbed is larger than the BH Schwarzchild radius Eq.~\ref{eq:bhunruh} holds  \cite{Giffin:2021kgb};  otherwise, Eq.~\ref{eq:baraccrate} dictates this rate. 
We find that DM accretion is very sub-leading and does not affect our final results.  
Additionally, the BH mass evolution is also governed by the rate of energy loss, characterized by Hawking evaporation, leading to
\begin{equation}\label{eq:bhevol}
    \frac{d\MBH}{dt}=\frac{d\MBH}{dt}\bigg|_{\rm Acc}-\frac{d\MBH}{dt}\bigg|_{\rm H}\,,
\end{equation}
with the  second term corresponding to the Hawking evaporation rate given by
\begin{align}
    \frac{d\MBH}{dt}\bigg|_{\rm H} = 5.34 \times 10^{16}\,f(\MBH)\, \left(\frac{{\rm kg}}{\MBH}\right)^2\,{\rm kg/s}\,,
\end{align}
where $f(\MBH)$ is a factor that takes into consideration the number of degrees of freedom emitted at a given BH temperature, $T_{\rm BH} \simeq 1.06 \times 10^{10}\,({\rm kg}/\MBH)$. Note that the standard treatment of assuming a blackbody radiation for BH evaporation is not accurate; instead, the factor needs to be calculated from the semi-analytic formula given in Ref.~\cite{MacGibbon:1991tj} or, alternatively, from the public code \texttt{BlackHawk} \cite{Arbey:2019mbc}. We use the semi-analytic formula for obtaining this factor and show the evaporation rate as a light blue line in Fig.~\ref{fig:accrates}. The function $f$ is normalized to unity for $\MBH \gg 10^{14}$ kg when the BH emits only photons, neutrinos (and antineutrinos), and $f = 15.35$ for $\MBH < 10^7$ kg, when the BH temperature is high enough for all SM particles to be emitted. The kink corresponds to the QCD confinement scale $\Lambda_{\rm QCD} \sim 250-300$ MeV, above which free rather than composite particles are emitted (see Refs.~\cite{Carr:2020gox,Ray:2023auh} for more details).  

As the two terms on the rhs of Eq.~\ref{eq:bhevol} have opposite dependence on the initial BH mass, equating the two gives us the critical mass $M_{\rm BH}^{\rm crit}$, above (below) which the BH will survive (evaporate). The critical mass corresponding to our benchmark NSs obtained after numerically solving Eq.~\ref{eq:bhevol} are shown in Table~\ref{tab:bhlimits}.
The value of $M_{\rm BH}^{\rm crit}$ is denoted by the gray dot-dashed line in Fig.~\ref{fig:accrates} for QMC-2. For an initial BH mass $M_\text{BH}^0 > \MBH^{\rm crit}$, the NS lifetime after the BH is formed  can be determined numerically by integrating Eq.~\ref{eq:bhevol}, 
\begin{align}
    \tau (M_\text{BH}^0) = \int_{M_\text{BH}^0}^{M_{\star}} d\MBH \left(\frac{d\MBH}{dt}\bigg|_{\rm Acc}-\frac{d\MBH}{dt}\bigg|_{\rm H}\right)^{-1}\,.
\end{align}
We find that the typical NS lifetime is $\mathcal{O}(1-100 {\rm~Myr})$ for $\MBH^{\rm crit} < M_\text{BH}^0 < \MBH^\text{U}$, in agreement with the numerical results of Ref.~\cite{Giffin:2021kgb}. 
\\

\noindent \textbf{Suppressed evaporation:} These limits have been derived under the assumption of standard Hawking evaporation. Hence, the limits are subject to change depending on the deviation from the standard treatment. First, the evaporation rate of BHs in the interior of a NS can be suppressed due to Pauli blocking of the emitted fermionic modes. However, these effects change the value of $\MBH^{\rm crit}$ by less than $10\%$~\cite{Autzen:2014tza}. Second, it has recently been shown that there can be a deviation from the standard treatment of Hawking evaporation due to the so-called ``memory burden effect'', implying that a BH possesses memory due to quantum effects, and that the evaporation rate once the BH has lost a fraction of its mass is suppressed \cite{Dvali:2020wft,Dvali:2024hsb}. This suppressed evaporation can thus extend the limits on DM masses that currently exist in literature by a factor of $100$~\cite{Basumatary:2024uwo}.  

In order to illustrate the effects of suppressed evaporation, let us just consider the accretion term in Eq.~\ref{eq:bhevol}, from which one can derive a limit on the initial BH mass below which the accretion timescale is always larger than the age of the star, and thus the NS survives. However, this requires a careful treatment. Initially, when $\MBH < \MBH^\text{U}$, the accretion term is given by Eq.~\ref{eq:bhunruh}. As the BH grows and $\MBH$ becomes larger than $\MBH^\text{U}$, the accretion term is then governed by Eq.~\ref{eq:baraccrate}. It can be seen in Fig.~\ref{fig:accrates} that the difference between the two rates is much  larger for heavy BH masses. Therefore, we solve the differential equation numerically taking both these accretion rates into account, i.e., the total accretion timescale is given by
\begin{equation}\label{eq:acctotal}
    \tau = \int_{\MBH^0}^{\MBH^\text{U}} d\MBH \frac{d\MBH}{dt}\bigg|_{\rm U} + \int_{\MBH^\text{U}}^{M_\star} d\MBH \frac{d\MBH}{dt}\bigg|_{\rm A} \,.
\end{equation}
This limit on $\MBH^0$ is given in the third column of Table.~\ref{tab:bhlimits} for $t_\star = 10$ Gyr. 
\begin{table}[!t]
    \centering
    \begin{tabular}{|c|c|c|c|}
        \hline
         \textbf{EoS} & $\MBH^\text{U}$ [kg] & $M_{\rm BH}^{\rm crit}~[{\rm kg}]$ & {$M_{\rm BH}^{\rm acc}~[{\rm kg}]$}\\
         \hline
         \textbf{QMC-1} & $2.06 \times 10^{12}$ & $1.27 \times 10^{11}$ & $1.07 \times 10^{10}$ \\
         \textbf{QMC-2} & $1.87 \times 10^{12}$&$1.28 \times 10^{11}$ & $1.07 \times 10^{10}$ \\
         \textbf{QMC-3} & $1.88 \times 10^{12}$ &$1.50 \times 10^{11}$ & $1.47 \times 10^{10}$ \\
         \hline         
    \end{tabular}
    \caption{Limits on the BH mass for the three benchmark NSs with EoS QMC. The first column denotes the Unruh limit, below which quantum effects are important. For $M_{\rm BH} < M_{\rm BH}^{\rm crit}$, the BH evaporates and no constraints can be derived from BH formation. The accretion limit in the third column denotes the BH mass, below which the NS will always survive, as the classical accretion timescale is larger than $t_\star = 10$ Gyr.}
    \label{tab:bhlimits}
\end{table}
Our limit on $\MBH$ assuming no evaporation ($\MBH^{\rm acc}$) shows that limits on BH masses can be at most be relaxed only by an order of magnitude compared to $\MBH^{\rm crit}$. Therefore, the limits from including the memory burden effect should lie in between the last two limits shown in Table~\ref{tab:bhlimits}. 

\section{Results}
\label{sec:results}

Previous work~\cite{McDermott:2011jp,Bell:2013xk,Dasgupta:2020dik,Lu:2024kiz,Basumatary:2024uwo} made  use of observations of the recycled MSP J0437-4715~\cite{Johnston:1993} to derive bounds on ADM accumulating in this pulsar. PSR J0437-4715 is the closest and brightest observed millisecond pulsar, as such it is not only rotation powered but, more importantly, it is in a binary system with a helium white dwarf~\cite{Bailyn:1993}. Here, it is worth remarking that the number of DM particles accumulated in a NS depends on the capture rate, whose expression in Eq.~\ref{eq:capturethin} does not account for the effect of a stellar companion. 
It is also believed that PSR J0437-4715 possesses a weak magnetic field with a complex structure~\cite{Choudhury:2024xbk}. Furthermore, the fact that we are dealing with a MSP, means that passive cooling models such as that of Fig.~\ref{fig:cool} that neither consider rotation, accretion from a binary  companion, nor magnetic fields are not suitable to infer core temperatures. MSPs, such as PSR J0437-4715, with  surface temperatures ${\cal O}(10^5\K)$ and characteristic ages ${\cal O}(\text{Gyr})$ are in agreement with internal reheating mechanisms due to rotation, occurring at late stages~\cite{Gonzalez:2010,Kantor:2021vwj}.

One may think then of using observations of isolated MSPs, such as PSR J2124-3358~\cite{Bailes:1997}  a nearby (at $410$~pc~\cite{Reardon:2015kba}), isolated MSP with a characteristic age of $3.8\Gyr$~\cite{Reardon:2015kba}. Despite the fact that it is isolated, there are still issues that prevent us from using this kind of observation. First, precisely because the MSP is isolated, there is no estimation for either its gravitational mass or its radius. 
Second, most MSPs are found in binary systems. In fact, the most massive progenitor  in the system evolves faster and becomes a normal pulsar until its radio emission shuts off. Later on, the pulsar is believed to accrete matter from the companion star, thereby acquiring angular momentum and becoming a MSP. This is why they are dubbed recycled pulsars. 
Thus,  isolated MSPs are thought to have lost their companion in a second supernova event or by ablation of the companion. 
Here, it is worth reminding the reader that pulsar characteristic ages (the quantity used to estimate pulsar ages in analysis similar to this work) are derived under the assumptions of (i) the magnetic field is well approximated by a dipole, (ii) the magnetic field does not decay with time, and (iii) the pulsar initial period was much smaller than  currently measured. In some cases, a braking index ($n=3$) is also assumed. These assumptions are very hard to meet by MSPs, in particular, that of having a constant magnetic field since it decays significantly with time. 
For this reason, we do not derive bounds from observed MSPs. 

\begin{figure}
    \centering
    \includegraphics[width=\columnwidth]{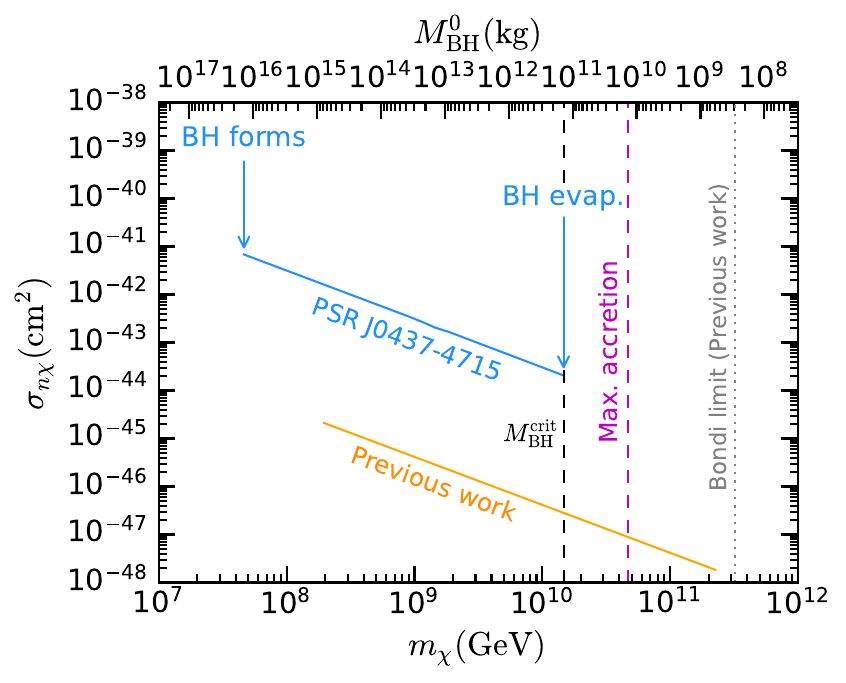}
    \caption{Comparison with previous work~\cite{Basumatary:2024uwo} for PSR J0437-4715. The  black dashed line represents the DM mass above which the BH always evaporates, and the NS will survive ($M_\text{BH}^\text{crit}$). The magenta dashed line denotes the maximum accretion limit in the absence of BH evaporation (the equivalent for this in the previous work is the Bondi limit depicted as a dotted gray line). For initial BH masses below this limit, the BH accretion timescale to consume the NS is larger than  $t_\star$.}
    \label{fig:PSRJ0437}
\end{figure}

However, for comparison with previous work~\cite{Basumatary:2024uwo} (orange line), we show in Fig.~\ref{fig:PSRJ0437} our result for PSR J0437-4715, whose characteristic age $6.7\pm0.2\Gyr$~\cite{Durant:2012,Gonzalez-Caniulef:2019wzi} is consistent with that of its white dwarf companion. The light blue line represents the region of the parameter space where a black hole forms at the NS center which eventually destroys the star. Note that the timescale for this to occur is much smaller than the age of the NS. For large DM masses ${\cal O}(10^{10}\GeV)$ the BH is expected to evaporate (dashed black line). Note the orders of magnitude difference in the DM-neutron scattering cross section from the previous result. This is due to the combined effect of a capture rate that properly accounts for the physics of NSs (see Fig.~\ref{fig:Csigma}), improved BH accretion rates in the classical and quantum regimes, and considering a more realistic DM distribution inside the NS which gives rise to Eq.~\ref{eq:nch}. The difference in the low DM mass end point is due again to an improved capture rate (also see the discussion below), and that in the large DM mass end point is due to the fact that we are not considering memory burden effects. Even so, we could only extend our result up to the magenta dashed line, which corresponds to the BH accretion timescale being always larger than the age of the NS. This maximum accretion  threshold is an order of magnitude lower than that found using Bondi accretion, Eq.~\ref{eq:Bondi}, (dotted gray line) in Ref.~\cite{Basumatary:2024uwo}. This difference is attributed to the different prescription for evaluating this limit, as highlighted in Eq.~\ref{eq:acctotal}, which accounts for quantum effects and a proper accretion rate in the classical regime.

\begin{figure}
    \centering
    \includegraphics[width=\columnwidth]{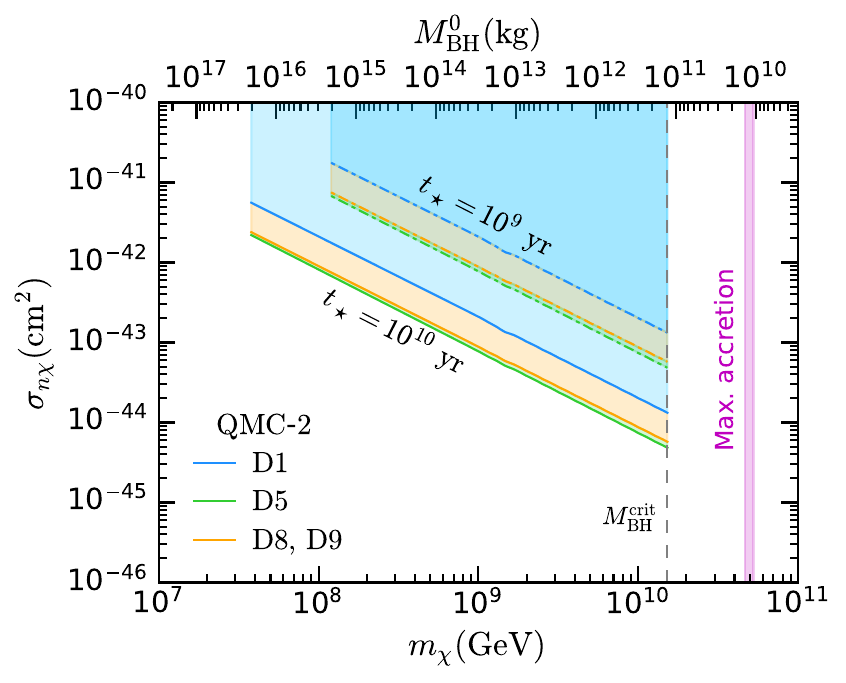}    
    \includegraphics[width=\columnwidth]{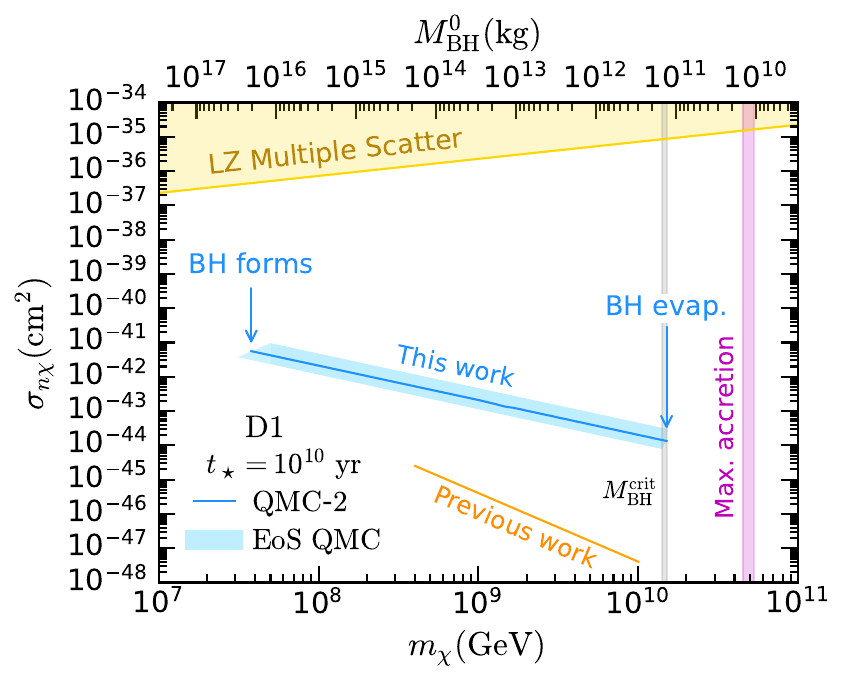}
    \caption{ 
    Top: Exclusion limits in the $\sigma_{n\chi}-m_\chi$ plane from NS destruction by black hole formation for scattering off neutrons for QMC-2 and operators D1, D5, D8 and D9, as well as $t_\star =1, 10$ Gyr.  
    Bottom: Same as before for operator D1 for $t_\star =10$ Gyr, considering a NS with EoS QMC (light blue band). 
    For comparison we show bounds from direct detection in the multiple scattering regime~\cite{LZ:2024psa} and results from  previous work~\cite{Garani:2018kkd}. 
    The gray dashed line (top) and the gray band (bottom)  represent the DM mass above which the BH always evaporates ($M_\text{BH}^\text{crit}$), whereas the magenta band represents the DM mass above which the accretion timescale is larger than the age of the star, and the NS will survive ($\MBH^{\rm acc}$).}
    \label{fig:r10}
\end{figure}

Given the aforementioned uncertainties in the estimation of the ages of observed pulsars, we then consider two benchmark NS ages: $t_\star = 1, 10~{\rm Gyr}$, along with our benchmark NSs with EoS QMC discussed in Section~\ref{sec:NSs}. As the NS lifetime (destruction time) for $M_\text{BH}^0 > M_\text{BH}^\text{crit}$ is much smaller than the age of the NS, we consider the DM capture over the NS age to determine possible constraints. Using the capture rates discussed above, one can determine first the parameter $\Lambda_q$ and then the low-energy cross section corresponding to the particular interaction $\sigma_{n\chi}^{c}$ using the expressions given in Ref.~\cite{Bell:2018pkk}, for a given $m_\chi$ for which $N_\chi (t_\star) = N_{\rm Ch}$, see Eq.~\ref{eq:nch}.

For $\sigma_{n\chi} > \sigma_{n\chi}^c$, the captured DM will lead to the formation of a BH in the interior of the neutron star with initial mass $M_\text{BH}^0 = m_\chi N_{\chi}(t_\star)$. This is shown by the shaded regions in the top panel of  Fig.~\ref{fig:r10} for operators D1, D5, D8 and D9, taking neutron targets,\footnote{The low-energy differential cross section for all ten  operators can be found in the Appendix of Ref.~\cite{Bell:2018pkk}.} QMC-2 ($1.5\Msun$) and $t_\star =1, 10$ Gyr. Further, as the capture rate cannot exceed the geometric value, we can determine the value of $m_\chi$, below which the total number of captured DM particles
does not reach the Chandrasekhar limit to form a BH within the lifetime of the NS. This is  the left unshaded region. It can be seen that the lowest DM mass that can be probed via NS is also an order an magnitude smaller than the previous results, where the end point is obtained with either $\sigma_{n\chi} \simeq 2 \times 10^{-45}$ or its multiple scattering equivalent, as the onset of geometric capture. As shown in Fig.~\ref{fig:Csigma}, the actual DM-neutron  threshold cross section is larger than this value by more than one order of magnitude than in previous calculations, a consequence of a suppressed capture rate. This is enough to counterbalance the fact that more DM particles are required in our case to form a BH than in previous work and even reduces the minimum DM mass to achieve this. 

In the right unshaded region, $M_\text{BH}^0 < \MBH^{\rm crit}$, and the BH formed will always evaporate, as discussed above. The value of $m_\chi$, for which $M_\text{BH}^0 = \MBH^{\rm crit}$, is depicted by the gray dashed line. We also show the value of $m_\chi$, corresponding to $M_\text{BH}^0 = \MBH^{\rm acc}$ (magenta dashed line), above which the BH accretion timescale is always larger than the age of NS. The end points of the band correspond to $t_\star=1$ and $10\Gyr$, respectively.

\begin{figure}[t]
    \centering
    \includegraphics[width=\columnwidth]{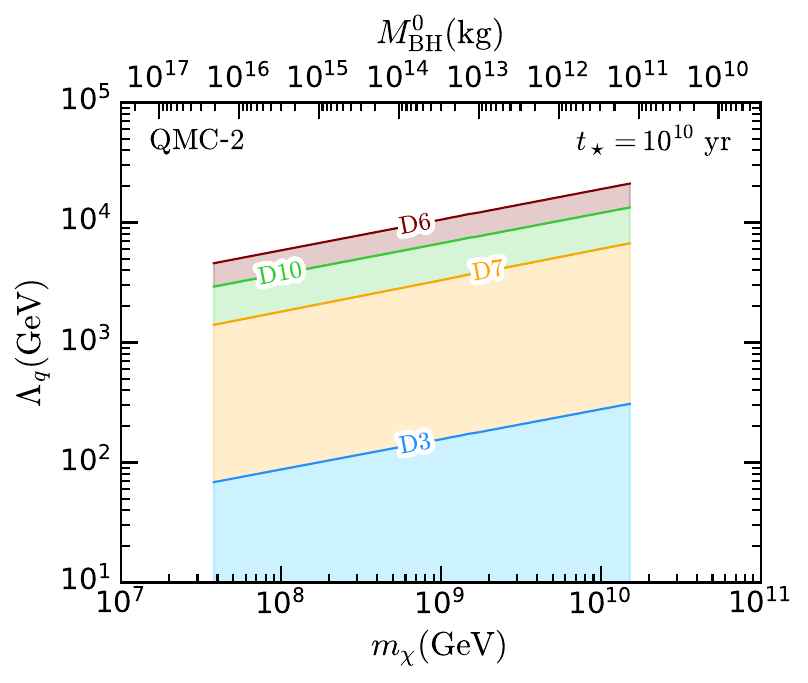}
    \caption{Same as Fig.~\ref{fig:r10}, but in terms of the parameter $\Lambda_q$ for some of the velocity and momentum suppressed EFT operators.}
    \label{fig:exlam}
\end{figure}

Our results show that possible bounds on DM-neutron scattering cross sections are relaxed by $2-3$ orders of magnitude from the existing calculations~\cite{Garani:2018kkd}, see bottom panel. 
This can be seen as a consequence of a suppressed captured rate as shown in Fig.~\ref{fig:Csigma}, as well as due to the modified Chandrasekhar limit in Eq.~\ref{eq:nch}, pushing the region toward larger values of $\sigma_{n\chi}$. 
The limits on the critical DM mass above which the formed BH evaporates is $m_\chi \gtrsim 10^{10}~{\rm GeV}$, similar to the previous results. However, due to the improved DM capture rate, the minimum DM mass below which one cannot reach the Chandrasekhar limit is an order of magnitude smaller than the previous one. We find that exclusion limits can thus be placed for $m_\chi > 10^7$ GeV.
In the bottom panel of Fig.~\ref{fig:r10}, we also show how our results vary with different NS compactness with QMC-1 ($1\Msun$) and QMC-3 ($1.9\Msun$) being less and more constraining in cross section, respectively. Since our results require DM capture in the multiple scattering regime, for comparison we show in yellow the region excluded by LZ searches for ultra heavy dark matter in this regime~\cite{LZ:2024psa}.

Finally, for completeness we show in Fig.~\ref{fig:exlam} the exclusion limits on the parameter $\Lambda_q$ for  operators not shown in Fig.~\ref{fig:r10},  namely,  D3, D6, D7, and D10. These are operators that in direct detection experiments are velocity or momentum suppressed. Therefore, $\Lambda_q$ cannot be easily converted into a non-relativistic scattering cross section. 
Note that these exclusion limits (region below the solid lines) have been derived by taking into account the scattering with all baryonic targets corresponding, which for the NS benchmark QMC-2 are neutrons and protons. 
For the remaining operators, D2 and D4, no meaningful constraints on $\Lambda_q$ from fermionic ADM triggering NS destruction can be derived.


\section{Conclusions}
\label{sec:conclusions}

Neutron stars (NSs) are promising probes for asymmetric dark matter (DM) models. In this work, we have revisited the constraints placed on fermionic asymmetric DM (ADM) from capture to collapse to a black hole (BH) in NSs. For the first time, 
we consistently use a relativistic equation of state to parametrize the microphysics relevant to each of the processes involved, namely, capture, thermalization,  self-gravitation,  collapse to a BH, and later BH accretion and evaporation.   Thus, all the effects relevant to the physics of neutron stars, such as general relativity effects and  Pauli blocking have been properly accounted for. 

Further, we have considered the effect of NS temperature evolution (cooling) on the thermalization timescale, which may lead to a more efficient thermalization of captured DM in young NSs. We find that, for fermionic ADM, the region of the parameter space of the model where constraints from BH formation can be placed all the captured DM can fully thermalize. 

We have reviewed the derivation of the Chandrasekhar limit for BH formation by ADM accumulation using a more realistic ADM distribution within the NS. This result differs from previous results by an order of magnitude more ADM particles required to trigger collapse to a BH. In addition, the fact that realistic ADM capture rates are suppressed, with respect to calculations that do not account for momentum-dependent nucleon couplings and nucleon strong interactions, 
shifts the minimum DM mass for BH formation  toward lower  $m_\chi$ values, meaning that lighter ADM than previously considered can be probed with NSs.

For BH accretion of NS matter, we have taken into account  
quantum effects that arise when the Schwarzschild radius of the BH is larger than the size of the nucleons. For classical accretion, we find that the commonly used Bondi accretion rate does not apply to this particular case,  since the conditions under which this rate was derived do not hold in the NS interior. 
This modifies the limit on $m_\chi$ above which no constraints from NS destruction can be placed, since the BH formed would evaporate. Finally, due to the uncertainty associated with estimating the age of old NSs that prevent us from using pulsar observations, we considered benchmark ages of $\mathcal{O}(\text{Gyr})$ to determine potential limits. 

While the original ADM hypothesis points toward $\mathcal{O}(\text{GeV})$ DM masses, ADM models can have larger (smaller) DM masses provided that the initial asymmetry in the dark sector is smaller (larger). 
In the case of fermionic ADM, we find that potential constraints from NS destruction are relevant only for DM masses above $10^7$ GeV. This showcases that NSs, apart from being complementary to direct detection experiments, have better sensitivity to heavier DM masses. 

To conclude, we find that, combining all the aforementioned  relevant effects, the existing bounds on fermionic ADM can be relaxed by orders of magnitude. Our analysis can easily be extended to the case of bosonic ADM, with a few exceptions, as there are some qualitative differences due to Bose-Einstein condensation. We leave this analysis for a future work.

\begin{acknowledgments}
We thank Joe Bramante, Raghuveer Garani, Chris Kouvaris,  Jose Pons, Nirmal Raj and Anupam Ray for helpful and insightful discussions. 
SR  was supported by the Fermi National Accelerator Laboratory (Fermilab), a U.S.
Department of Energy, Office of Science, HEP User Facility. 
SR  acknowledges CERN TH Department for its hospitality while this research was being carried out. DV is supported by the “Generalitat Valenciana” through the
GenT Excellence Program (CIDEGENT/2020/020), and greatly acknowledges the hospitality of  Fermilab's Theory Division where this project was initiated. 
GB was supported by the Australian Research Council through the ARC Centre of Excellence for Dark Matter Particle Physics, CE200100008. 
This project has received funding from the European Union’s Horizon Europe research and innovation programme under the Marie Skłodowska-Curie Staff Exchange grant agreement No 101086085 - ASYMMETRY. 

\end{acknowledgments}

\bibliographystyle{utphys}
\bibliography{refs}

\providecommand{\href}[2]{#2}\begingroup\raggedright\begin{thebibliography}{10}

\bibitem{Kouvaris:2007ay}
C.~Kouvaris, ``{WIMP Annihilation and Cooling of Neutron Stars},'' \href{http://dx.doi.org/10.1103/PhysRevD.77.023006}{{\em Phys. Rev. D} {\bfseries 77} (2008) 023006}, \href{http://arxiv.org/abs/0708.2362}{{\ttfamily arXiv:0708.2362 [astro-ph]}}.

\bibitem{Bertone:2007ae}
G.~Bertone and M.~Fairbairn, ``{Compact Stars as Dark Matter Probes},'' \href{http://dx.doi.org/10.1103/PhysRevD.77.043515}{{\em Phys. Rev. D} {\bfseries 77} (2008) 043515}, \href{http://arxiv.org/abs/0709.1485}{{\ttfamily arXiv:0709.1485 [astro-ph]}}.

\bibitem{Kouvaris:2010vv}
C.~Kouvaris and P.~Tinyakov, ``{Can Neutron stars constrain Dark Matter?},'' \href{http://dx.doi.org/10.1103/PhysRevD.82.063531}{{\em Phys. Rev. D} {\bfseries 82} (2010) 063531}, \href{http://arxiv.org/abs/1004.0586}{{\ttfamily arXiv:1004.0586 [astro-ph.GA]}}.

\bibitem{Kaplan:2009ag}
D.~E. Kaplan, M.~A. Luty, and K.~M. Zurek, ``{Asymmetric Dark Matter},'' \href{http://dx.doi.org/10.1103/PhysRevD.79.115016}{{\em Phys. Rev. D} {\bfseries 79} (2009) 115016}, \href{http://arxiv.org/abs/0901.4117}{{\ttfamily arXiv:0901.4117 [hep-ph]}}.

\bibitem{Petraki:2013wwa}
K.~Petraki and R.~R. Volkas, ``{Review of asymmetric dark matter},'' \href{http://dx.doi.org/10.1142/S0217751X13300287}{{\em Int. J. Mod. Phys. A} {\bfseries 28} (2013) 1330028}, \href{http://arxiv.org/abs/1305.4939}{{\ttfamily arXiv:1305.4939 [hep-ph]}}.

\bibitem{Zurek:2013wia}
K.~M. Zurek, ``{Asymmetric Dark Matter: Theories, Signatures, and Constraints},'' \href{http://dx.doi.org/10.1016/j.physrep.2013.12.001}{{\em Phys. Rept.} {\bfseries 537} (2014) 91--121}, \href{http://arxiv.org/abs/1308.0338}{{\ttfamily arXiv:1308.0338 [hep-ph]}}.

\bibitem{Goldman:1989nd}
I.~Goldman and S.~Nussinov, ``{Weakly Interacting Massive Particles and Neutron Stars},''
\href{http://dx.doi.org/10.1103/PhysRevD.40.3221}{{\em Phys. Rev.} {\bfseries D40} (1989) 3221--3230}.

\bibitem{Kouvaris:2010jy}
C.~Kouvaris and P.~Tinyakov, ``{Constraining Asymmetric Dark Matter through observations of compact stars},'' \href{http://dx.doi.org/10.1103/PhysRevD.83.083512}{{\em Phys. Rev.} {\bfseries D83} (2011) 083512},
\href{http://arxiv.org/abs/1012.2039}{{\ttfamily arXiv:1012.2039 [astro-ph.HE]}}.

\bibitem{McDermott:2011jp}
S.~D. McDermott, H.-B. Yu, and K.~M. Zurek, ``{Constraints on Scalar Asymmetric Dark Matter from Black Hole Formation in Neutron Stars},'' \href{http://dx.doi.org/10.1103/PhysRevD.85.023519}{{\em Phys. Rev.} {\bfseries D85} (2012) 023519},
\href{http://arxiv.org/abs/1103.5472}{{\ttfamily arXiv:1103.5472 [hep-ph]}}.

\bibitem{Kouvaris:2011fi}
C.~Kouvaris and P.~Tinyakov, ``{Excluding Light Asymmetric Bosonic Dark Matter},'' \href{http://dx.doi.org/10.1103/PhysRevLett.107.091301}{{\em Phys. Rev. Lett.} {\bfseries 107} (2011) 091301},
\href{http://arxiv.org/abs/1104.0382}{{\ttfamily arXiv:1104.0382 [astro-ph.CO]}}.

\bibitem{Guver:2012ba}
T.~Güver, A.~E. Erkoca, M.~Hall~Reno, and I.~Sarcevic, ``{On the capture of dark matter by neutron stars},'' \href{http://dx.doi.org/10.1088/1475-7516/2014/05/013}{{\em JCAP} {\bfseries 1405} (2014) 013},
\href{http://arxiv.org/abs/1201.2400}{{\ttfamily arXiv:1201.2400 [hep-ph]}}.

\bibitem{Kouvaris:2012dz}
C.~Kouvaris and P.~Tinyakov, ``{(Not)-constraining heavy asymmetric bosonic dark matter},'' \href{http://dx.doi.org/10.1103/PhysRevD.87.123537}{{\em Phys. Rev. D} {\bfseries 87} no.~12, (2013) 123537}, \href{http://arxiv.org/abs/1212.4075}{{\ttfamily arXiv:1212.4075 [astro-ph.HE]}}.

\bibitem{Bell:2013xk}
N.~F. Bell, A.~Melatos, and K.~Petraki, ``{Realistic neutron star constraints on bosonic asymmetric dark matter},'' \href{http://dx.doi.org/10.1103/PhysRevD.87.123507}{{\em Phys. Rev.} {\bfseries D87} no.~12, (2013) 123507},
\href{http://arxiv.org/abs/1301.6811}{{\ttfamily arXiv:1301.6811 [hep-ph]}}.

\bibitem{Bramante:2013nma}
J.~Bramante, K.~Fukushima, J.~Kumar, and E.~Stopnitzky, ``{Bounds on self-interacting fermion dark matter from observations of old neutron stars},'' \href{http://dx.doi.org/10.1103/PhysRevD.89.015010}{{\em Phys. Rev.} {\bfseries D89} no.~1, (2014) 015010},
\href{http://arxiv.org/abs/1310.3509}{{\ttfamily arXiv:1310.3509 [hep-ph]}}.

\bibitem{Bramante:2014zca}
J.~Bramante and T.~Linden, ``{Detecting Dark Matter with Imploding Pulsars in the Galactic Center},'' \href{http://dx.doi.org/10.1103/PhysRevLett.113.191301}{{\em Phys. Rev. Lett.} {\bfseries 113} no.~19, (2014) 191301}, \href{http://arxiv.org/abs/1405.1031}{{\ttfamily arXiv:1405.1031 [astro-ph.HE]}}.

\bibitem{Bramante:2015dfa}
J.~Bramante and F.~Elahi, ``{Higgs portals to pulsar collapse},'' \href{http://dx.doi.org/10.1103/PhysRevD.91.115001}{{\em Phys. Rev. D} {\bfseries 91} no.~11, (2015) 115001}, \href{http://arxiv.org/abs/1504.04019}{{\ttfamily arXiv:1504.04019 [hep-ph]}}.

\bibitem{Garani:2018kkd}
R.~Garani, Y.~Genolini, and T.~Hambye, ``{New Analysis of Neutron Star Constraints on Asymmetric Dark Matter},'' \href{http://dx.doi.org/10.1088/1475-7516/2019/05/035}{{\em JCAP} {\bfseries 05} (2019) 035}, \href{http://arxiv.org/abs/1812.08773}{{\ttfamily arXiv:1812.08773 [hep-ph]}}.

\bibitem{Dasgupta:2020dik}
B.~Dasgupta, A.~Gupta, and A.~Ray, ``{Dark matter capture in celestial objects: light mediators, self-interactions, and complementarity with direct detection},'' \href{http://dx.doi.org/10.1088/1475-7516/2020/10/023}{{\em JCAP} {\bfseries 10} (2020) 023}, \href{http://arxiv.org/abs/2006.10773}{{\ttfamily arXiv:2006.10773 [hep-ph]}}.

\bibitem{Giffin:2021kgb}
P.~Giffin, J.~Lloyd, S.~D. McDermott, and S.~Profumo, ``{Neutron star quantum death by small black holes},'' \href{http://dx.doi.org/10.1103/PhysRevD.105.123030}{{\em Phys. Rev. D} {\bfseries 105} no.~12, (2022) 123030}, \href{http://arxiv.org/abs/2105.06504}{{\ttfamily arXiv:2105.06504 [hep-ph]}}.

\bibitem{Lu:2024kiz}
C.-T. Lu, A.~K. Mishra, and L.~Wu, ``{Constraining bosonic dark matter-baryon interactions from neutron star collapse},'' \href{http://dx.doi.org/10.1088/1475-7516/2024/09/066}{{\em JCAP} {\bfseries 09} (2024) 066}, \href{http://arxiv.org/abs/2404.07187}{{\ttfamily arXiv:2404.07187 [hep-ph]}}.

\bibitem{Liu:2024qbe}
N.~Liu and A.~K. Mishra, ``{Neutron star collapse from accretion: A probe of massive dark matter particles},'' \href{http://dx.doi.org/10.1016/j.dark.2024.101740}{{\em Phys. Dark Univ.} {\bfseries 47} (2025) 101740}, \href{http://arxiv.org/abs/2408.00594}{{\ttfamily arXiv:2408.00594 [astro-ph.CO]}}.

\bibitem{Dutta:2024vzw}
K.~Dutta, D.~Ghosh, and B.~Mukhopadhyaya, ``{Improved treatment of bosonic dark matter dynamics in neutron stars: consequences and constraints},'' \href{http://dx.doi.org/10.1088/1475-7516/2024/12/053}{{\em JCAP} {\bfseries 12} (2024) 053}, \href{http://arxiv.org/abs/2408.16091}{{\ttfamily arXiv:2408.16091 [hep-ph]}}.

\bibitem{Basumatary:2024uwo}
U.~Basumatary, N.~Raj, and A.~Ray, ``{Beyond Hawking evaporation of black holes formed by dark matter in compact stars},'' \href{http://dx.doi.org/10.1103/PhysRevD.111.L041306}{{\em Phys. Rev. D} {\bfseries 111} no.~4, (2025) L041306}, \href{http://arxiv.org/abs/2410.22702}{{\ttfamily arXiv:2410.22702 [hep-ph]}}.

\bibitem{Bell:2020jou}
N.~F. Bell, G.~Busoni, S.~Robles, and M.~Virgato, ``{Improved Treatment of Dark Matter Capture in Neutron Stars},'' \href{http://dx.doi.org/10.1088/1475-7516/2020/09/028}{{\em JCAP} {\bfseries 09} (2020) 028}, \href{http://arxiv.org/abs/2004.14888}{{\ttfamily arXiv:2004.14888 [hep-ph]}}.

\bibitem{Bell:2020lmm}
N.~F. Bell, G.~Busoni, S.~Robles, and M.~Virgato, ``{Improved Treatment of Dark Matter Capture in Neutron Stars II: Leptonic Targets},'' \href{http://dx.doi.org/10.1088/1475-7516/2021/03/086}{{\em JCAP} {\bfseries 03} (2021) 086}, \href{http://arxiv.org/abs/2010.13257}{{\ttfamily arXiv:2010.13257 [hep-ph]}}.

\bibitem{Bell:2020obw}
N.~F. Bell, G.~Busoni, T.~F. Motta, S.~Robles, A.~W. Thomas, and M.~Virgato, ``{Nucleon Structure and Strong Interactions in Dark Matter Capture in Neutron Stars},'' \href{http://dx.doi.org/10.1103/PhysRevLett.127.111803}{{\em Phys. Rev. Lett.} {\bfseries 127} no.~11, (2021) 111803}, \href{http://arxiv.org/abs/2012.08918}{{\ttfamily arXiv:2012.08918 [hep-ph]}}.

\bibitem{Anzuini:2021lnv}
F.~Anzuini, N.~F. Bell, G.~Busoni, T.~F. Motta, S.~Robles, A.~W. Thomas, and M.~Virgato, ``{Improved treatment of dark matter capture in neutron stars III: nucleon and exotic targets},'' \href{http://dx.doi.org/10.1088/1475-7516/2021/11/056}{{\em JCAP} {\bfseries 11} (2021) 056}, \href{http://arxiv.org/abs/2108.02525}{{\ttfamily arXiv:2108.02525 [hep-ph]}}. [Erratum: JCAP 04, E02 (2024)].

\bibitem{Bell:2023ysh}
N.~F. Bell, G.~Busoni, S.~Robles, and M.~Virgato, ``{Thermalization and annihilation of dark matter in neutron stars},'' \href{http://dx.doi.org/10.1088/1475-7516/2024/04/006}{{\em JCAP} {\bfseries 04} (2024) 006}, \href{http://arxiv.org/abs/2312.11892}{{\ttfamily arXiv:2312.11892 [hep-ph]}}.

\bibitem{Guichon:2018uew}
P.~Guichon, J.~Stone, and A.~Thomas, ``{Quark\textendash{}Meson-Coupling (QMC) model for finite nuclei, nuclear matter and beyond},'' \href{http://dx.doi.org/10.1016/j.ppnp.2018.01.008}{{\em Prog. Part. Nucl. Phys.} {\bfseries 100} (2018) 262--297}, \href{http://arxiv.org/abs/1802.08368}{{\ttfamily arXiv:1802.08368 [nucl-th]}}.

\bibitem{Motta:2019tjc}
T.~Motta, A.~Kalaitzis, S.~Anti\'c, P.~Guichon, J.~Stone, and A.~Thomas, ``{Isovector Effects in Neutron Stars, Radii and the GW170817 Constraint},'' \href{http://dx.doi.org/10.3847/1538-4357/ab218e}{{\em Astrophys. J.} {\bfseries 878} no.~2, (2019) 159}, \href{http://arxiv.org/abs/1904.03794}{{\ttfamily arXiv:1904.03794 [nucl-th]}}.

\bibitem{Tolman:1939jz}
R.~C. Tolman, ``{Static solutions of Einstein's field equations for spheres of fluid},''
\href{http://dx.doi.org/10.1103/PhysRev.55.364}{{\em Phys. Rev.} {\bfseries 55} (1939) 364--373}.

\bibitem{Oppenheimer:1939ne}
J.~R. Oppenheimer and G.~M. Volkoff, ``{On Massive neutron cores},''
\href{http://dx.doi.org/10.1103/PhysRev.55.374}{{\em Phys. Rev.} {\bfseries 55} (1939) 374--381}.

\bibitem{Ootes:2016oib}
L.~S. Ootes, D.~Page, R.~Wijnands, and N.~Degenaar, ``{Neutron star crust cooling in KS 1731{\ensuremath{-}}260: the influence of accretion outburst variability on the crustal temperature evolution},'' \href{http://dx.doi.org/10.1093/mnrasl/slw131}{{\em Mon. Not. Roy. Astron. Soc.} {\bfseries 461} no.~4, (2016) 4400--4405}, \href{http://arxiv.org/abs/1606.01923}{{\ttfamily arXiv:1606.01923 [astro-ph.HE]}}. [Erratum: Mon.Not.Roy.Astron.Soc. 485, L16 (2019)].

\bibitem{Nomoto:1987}
K.~{Nomoto} and S.~{Tsuruta}, ``{Cooling of Neutron Stars: Effects of the Finite Time Scale of Thermal Conduction},'' \href{http://dx.doi.org/10.1086/164914}{{\em \apj} {\bfseries 312} (Jan., 1987) 711}.

\bibitem{Reisenegger:1994be}
A.~Reisenegger, ``{Deviations from chemical equilibrium due to spindown as an internal heat source in neutron stars},'' \href{http://dx.doi.org/10.1086/175480}{{\em Astrophys. J.} {\bfseries 442} (1995) 749}, \href{http://arxiv.org/abs/astro-ph/9410035}{{\ttfamily arXiv:astro-ph/9410035}}.

\bibitem{Reisenegger:1997}
A.~{Reisenegger}, ``{Constraining Dense Matter Superfluidity through Thermal Emission from Millisecond Pulsars},'' \href{http://dx.doi.org/10.1086/304417}{{\em \apj} {\bfseries 485} no.~1, (Aug., 1997) 313--318}, \href{http://arxiv.org/abs/astro-ph/9612179}{{\ttfamily arXiv:astro-ph/9612179 [astro-ph]}}.

\bibitem{Fernandez:2005cg}
R.~Fernandez and A.~Reisenegger, ``{Rotochemical heating in millisecond pulsars. Formalism and non-superfluid case},'' \href{http://dx.doi.org/10.1086/429551}{{\em Astrophys. J.} {\bfseries 625} (2005) 291--306}, \href{http://arxiv.org/abs/astro-ph/0502116}{{\ttfamily arXiv:astro-ph/0502116}}.

\bibitem{Gonzalez:2010}
D.~{Gonzalez} and A.~{Reisenegger}, ``{Internal heating of old neutron stars: contrasting different mechanisms},'' \href{http://dx.doi.org/10.1051/0004-6361/201015084}{{\em Astron. Astrophys.} {\bfseries 522} (Nov., 2010) A16}, \href{http://arxiv.org/abs/1005.5699}{{\ttfamily arXiv:1005.5699 [astro-ph.HE]}}.

\bibitem{Alpar:1984}
M.~A. {Alpar}, D.~{Pines}, P.~W. {Anderson}, and J.~{Shaham}, ``{Vortex creep and the internal temperature of neutron stars. I - General theory},'' \href{http://dx.doi.org/10.1086/161616}{{\em \apj} {\bfseries 276} (Jan., 1984) 325--334}.

\bibitem{Shibazaki:1989}
N.~{Shibazaki} and F.~K. {Lamb}, ``{Neutron Star Evolution with Internal Heating},'' \href{http://dx.doi.org/10.1086/168062}{{\em \apj} {\bfseries 346} (Nov., 1989) 808}.

\bibitem{Larson:1998it}
M.~B. Larson and B.~Link, ``{Superfluid friction and late-time thermal evolution of neutron stars},'' \href{http://dx.doi.org/10.1086/307532}{{\em Astrophys. J.} {\bfseries 521} (1999) 271}, \href{http://arxiv.org/abs/astro-ph/9810441}{{\ttfamily arXiv:astro-ph/9810441}}.

\bibitem{Gusakov:2015kaa}
M.~E. Gusakov, E.~M. Kantor, and A.~Reisenegger, ``{Rotation-induced deep crustal heating of millisecond pulsars},'' \href{http://dx.doi.org/10.1093/mnrasl/slv095}{{\em Mon. Not. Roy. Astron. Soc.} {\bfseries 453} no.~1, (2015) L36--L40}, \href{http://arxiv.org/abs/1507.04586}{{\ttfamily arXiv:1507.04586 [astro-ph.HE]}}.

\bibitem{Haensel:1990}
P.~{Haensel}, V.~A. {Urpin}, and D.~G. {Iakovlev}, ``{Ohmic decay of internal magnetic fields in neutron stars},'' {\em Astron. Astrophys.} {\bfseries 229} no.~1, (Mar., 1990) 133--137.

\bibitem{Miralles:1998na}
J.~A. Miralles, V.~Urpin, and D.~Konenkov, ``{Joule heating and the thermal evolution of old neutron stars},'' \href{http://dx.doi.org/10.1086/305967}{{\em Astrophys. J.} {\bfseries 503} (1998) 368}, \href{http://arxiv.org/abs/astro-ph/9803063}{{\ttfamily arXiv:astro-ph/9803063}}.

\bibitem{Pons:2008fd}
J.~Pons, J.~Miralles, and U.~Geppert, ``{Magneto--thermal evolution of neutron stars},'' \href{http://dx.doi.org/10.1051/0004-6361:200811229}{{\em Astron. Astrophys.} {\bfseries 496} (2009) 207--216}, \href{http://arxiv.org/abs/0812.3018}{{\ttfamily arXiv:0812.3018 [astro-ph]}}.

\bibitem{Guillot:2019ugf}
S.~Guillot, G.~G. Pavlov, C.~Reyes, A.~Reisenegger, L.~Rodriguez, B.~Rangelov, and O.~Kargaltsev, ``{Hubble Space Telescope Nondetection of PSR J2144\textendash{}3933: The Coldest Known Neutron Star},'' \href{http://dx.doi.org/10.3847/1538-4357/ab0f38}{{\em Astrophys. J.} {\bfseries 874} no.~2, (2019) 175}, \href{http://arxiv.org/abs/1901.07998}{{\ttfamily arXiv:1901.07998 [astro-ph.HE]}}.

\bibitem{Bell:2018pkk}
N.~F. Bell, G.~Busoni, and S.~Robles, ``{Heating up Neutron Stars with Inelastic Dark Matter},'' \href{http://dx.doi.org/10.1088/1475-7516/2018/09/018}{{\em JCAP} {\bfseries 1809} no.~09, (2018) 018},
\href{http://arxiv.org/abs/1807.02840}{{\ttfamily arXiv:1807.02840 [hep-ph]}}.

\bibitem{Bramante:2017xlb}
J.~Bramante, A.~Delgado, and A.~Martin, ``{Multiscatter stellar capture of dark matter},'' \href{http://dx.doi.org/10.1103/PhysRevD.96.063002}{{\em Phys. Rev.} {\bfseries D96} no.~6, (2017) 063002},
\href{http://arxiv.org/abs/1703.04043}{{\ttfamily arXiv:1703.04043 [hep-ph]}}.

\bibitem{Joglekar:2020liw}
A.~Joglekar, N.~Raj, P.~Tanedo, and H.-B. Yu, ``{Dark kinetic heating of neutron stars from contact interactions with relativistic targets},'' \href{http://dx.doi.org/10.1103/PhysRevD.102.123002}{{\em Phys. Rev. D} {\bfseries 102} no.~12, (2020) 123002}, \href{http://arxiv.org/abs/2004.09539}{{\ttfamily arXiv:2004.09539 [hep-ph]}}.

\bibitem{Baryakhtar:2017dbj}
M.~Baryakhtar, J.~Bramante, S.~W. Li, T.~Linden, and N.~Raj, ``{Dark Kinetic Heating of Neutron Stars and An Infrared Window On WIMPs, SIMPs, and Pure Higgsinos},'' \href{http://dx.doi.org/10.1103/PhysRevLett.119.131801}{{\em Phys. Rev. Lett.} {\bfseries 119} no.~13, (2017) 131801},
\href{http://arxiv.org/abs/1704.01577}{{\ttfamily arXiv:1704.01577 [hep-ph]}}.

\bibitem{Bertoni:2013bsa}
B.~Bertoni, A.~E. Nelson, and S.~Reddy, ``{Dark Matter Thermalization in Neutron Stars},'' \href{http://dx.doi.org/10.1103/PhysRevD.88.123505}{{\em Phys. Rev.} {\bfseries D88} (2013) 123505},
\href{http://arxiv.org/abs/1309.1721}{{\ttfamily arXiv:1309.1721 [hep-ph]}}.

\bibitem{Steigerwald:2022pjo}
H.~Steigerwald, V.~Marra, and S.~Profumo, ``{Revisiting constraints on asymmetric dark matter from collapse in white dwarf stars},'' \href{http://dx.doi.org/10.1103/PhysRevD.105.083507}{{\em Phys. Rev. D} {\bfseries 105} no.~8, (2022) 083507}, \href{http://arxiv.org/abs/2203.09054}{{\ttfamily arXiv:2203.09054 [astro-ph.CO]}}.

\bibitem{Bell:2024qmj}
N.~F. Bell, G.~Busoni, S.~Robles, and M.~Virgato, ``{Heavy dark matter in white dwarfs: multiple-scattering capture and thermalization},'' \href{http://dx.doi.org/10.1088/1475-7516/2024/07/051}{{\em JCAP} {\bfseries 07} (2024) 051}, \href{http://arxiv.org/abs/2404.16272}{{\ttfamily arXiv:2404.16272 [hep-ph]}}.

\bibitem{Bondi:1952}
H.~{Bondi}, ``{On spherically symmetrical accretion},'' \href{http://dx.doi.org/10.1093/mnras/112.2.195}{{\em Mon. Not. R. Astron. Soc.} {\bfseries 112} (Jan., 1952) 195}.

\bibitem{Michel:1972oeq}
F.~C. Michel, ``{Accretion of matter by condensed objects},'' \href{http://dx.doi.org/10.1007/BF00649949}{{\em Astrophys. Space Sci.} {\bfseries 15} no.~1, (1972) 153--160}.

\bibitem{Shapiro:1983du}
S.~L. Shapiro and S.~A. Teukolsky, \href{http://dx.doi.org/10.1002/9783527617661}{{\em {Black holes, white dwarfs, and neutron stars: The physics of compact objects}}}.
\newblock Wiley-VCH, 1983.

\bibitem{Bondi:1944}
H.~{Bondi} and F.~{Hoyle}, ``{On the mechanism of accretion by stars},'' \href{http://dx.doi.org/10.1093/mnras/104.5.273}{{\em Mon. Not. R. Astron. Soc.} {\bfseries 104} (Jan., 1944) 273}.

\bibitem{Shima:1985}
E.~{Shima}, T.~{Matsuda}, H.~{Takeda}, and K.~{Sawada}, ``{Hydrodynamic calculations of axisymmetric accretion flow},'' \href{http://dx.doi.org/10.1093/mnras/217.2.367}{{\em Mon. Not. R. Astron. Soc.} {\bfseries 217} (Nov., 1985) 367--386}.

\bibitem{Edgar:2004mk}
R.~G. Edgar, ``{A Review of Bondi-Hoyle-Lyttleton accretion},'' \href{http://dx.doi.org/10.1016/j.newar.2004.06.001}{{\em New Astron. Rev.} {\bfseries 48} (2004) 843--859}, \href{http://arxiv.org/abs/astro-ph/0406166}{{\ttfamily arXiv:astro-ph/0406166}}.

\bibitem{East:2019dxt}
W.~E. East and L.~Lehner, ``{Fate of a neutron star with an endoparasitic black hole and implications for dark matter},'' \href{http://dx.doi.org/10.1103/PhysRevD.100.124026}{{\em Phys. Rev. D} {\bfseries 100} no.~12, (2019) 124026}, \href{http://arxiv.org/abs/1909.07968}{{\ttfamily arXiv:1909.07968 [gr-qc]}}.

\bibitem{Richards:2021upu}
C.~B. Richards, T.~W. Baumgarte, and S.~L. Shapiro, ``{Accretion onto a small black hole at the center of a neutron star},'' \href{http://dx.doi.org/10.1103/PhysRevD.103.104009}{{\em Phys. Rev. D} {\bfseries 103} no.~10, (2021) 104009}, \href{http://arxiv.org/abs/2102.09574}{{\ttfamily arXiv:2102.09574 [astro-ph.HE]}}.

\bibitem{Autzen:2014tza}
M.~Autzen and C.~Kouvaris, ``{Blocking the Hawking Radiation},'' \href{http://dx.doi.org/10.1103/PhysRevD.89.123519}{{\em Phys. Rev. D} {\bfseries 89} no.~12, (2014) 123519}, \href{http://arxiv.org/abs/1403.1072}{{\ttfamily arXiv:1403.1072 [astro-ph.SR]}}.

\bibitem{Unruh:1976fm}
W.~G. Unruh, ``{Absorption Cross-Section of Small Black Holes},'' \href{http://dx.doi.org/10.1103/PhysRevD.14.3251}{{\em Phys. Rev. D} {\bfseries 14} (1976) 3251--3259}.

\bibitem{Kouvaris:2013kra}
C.~Kouvaris and P.~Tinyakov, ``{Growth of Black Holes in the interior of Rotating Neutron Stars},'' \href{http://dx.doi.org/10.1103/PhysRevD.90.043512}{{\em Phys. Rev. D} {\bfseries 90} no.~4, (2014) 043512}, \href{http://arxiv.org/abs/1312.3764}{{\ttfamily arXiv:1312.3764 [astro-ph.SR]}}.

\bibitem{MacGibbon:1991tj}
J.~H. MacGibbon, ``{Quark and gluon jet emission from primordial black holes. 2. The Lifetime emission},'' \href{http://dx.doi.org/10.1103/PhysRevD.44.376}{{\em Phys. Rev. D} {\bfseries 44} (1991) 376--392}.

\bibitem{Arbey:2019mbc}
A.~Arbey and J.~Auffinger, ``{BlackHawk: A public code for calculating the Hawking evaporation spectra of any black hole distribution},'' \href{http://dx.doi.org/10.1140/epjc/s10052-019-7161-1}{{\em Eur. Phys. J. C} {\bfseries 79} no.~8, (2019) 693}, \href{http://arxiv.org/abs/1905.04268}{{\ttfamily arXiv:1905.04268 [gr-qc]}}.

\bibitem{Carr:2020gox}
B.~Carr, K.~Kohri, Y.~Sendouda, and J.~Yokoyama, ``{Constraints on primordial black holes},'' \href{http://dx.doi.org/10.1088/1361-6633/ac1e31}{{\em Rept. Prog. Phys.} {\bfseries 84} no.~11, (2021) 116902}, \href{http://arxiv.org/abs/2002.12778}{{\ttfamily arXiv:2002.12778 [astro-ph.CO]}}.

\bibitem{Ray:2023auh}
A.~Ray, ``{Celestial objects as strongly-interacting nonannihilating dark matter detectors},'' \href{http://dx.doi.org/10.1103/PhysRevD.107.083012}{{\em Phys. Rev. D} {\bfseries 107} no.~8, (2023) 083012}, \href{http://arxiv.org/abs/2301.03625}{{\ttfamily arXiv:2301.03625 [hep-ph]}}.

\bibitem{Dvali:2020wft}
G.~Dvali, L.~Eisemann, M.~Michel, and S.~Zell, ``{Black hole metamorphosis and stabilization by memory burden},'' \href{http://dx.doi.org/10.1103/PhysRevD.102.103523}{{\em Phys. Rev. D} {\bfseries 102} no.~10, (2020) 103523}, \href{http://arxiv.org/abs/2006.00011}{{\ttfamily arXiv:2006.00011 [hep-th]}}.

\bibitem{Dvali:2024hsb}
G.~Dvali, J.~S. Valbuena-Berm\'udez, and M.~Zantedeschi, ``{Memory burden effect in black holes and solitons: Implications for PBH},'' \href{http://dx.doi.org/10.1103/PhysRevD.110.056029}{{\em Phys. Rev. D} {\bfseries 110} no.~5, (2024) 056029}, \href{http://arxiv.org/abs/2405.13117}{{\ttfamily arXiv:2405.13117 [hep-th]}}.

\bibitem{Johnston:1993}
S.~{Johnston}, D.~R. {Lorimer}, P.~A. {Harrison}, M.~{Bailes}, A.~G. {Lynet}, J.~F. {Bell}, V.~M. {Kaspi}, R.~N. {Manchester}, N.~{D'Amico}, L.~{Nleastrol}, and J.~{Shengzhen}, ``{Discovery of a very bright, nearby binary millisecond pulsar},'' \href{http://dx.doi.org/10.1038/361613a0}{{\em Nature} {\bfseries 361} no.~6413, (Feb., 1993) 613--615}.

\bibitem{Bailyn:1993}
C.~D. {Bailyn}, ``{The Optical Counterpart of the Bright Nearby Millisecond Pulsar PSR J0437-4715},'' \href{http://dx.doi.org/10.1086/186918}{{\em Astrophys. J. Lett.} {\bfseries 411} (July, 1993) L83}.

\bibitem{Choudhury:2024xbk}
D.~Choudhury {\em et~al.}, ``{A NICER View of the Nearest and Brightest Millisecond Pulsar: PSR J0437\textendash{}4715},'' \href{http://dx.doi.org/10.3847/2041-8213/ad5a6f}{{\em Astrophys. J. Lett.} {\bfseries 971} no.~1, (2024) L20}, \href{http://arxiv.org/abs/2407.06789}{{\ttfamily arXiv:2407.06789 [astro-ph.HE]}}.

\bibitem{Kantor:2021vwj}
E.~M. Kantor and M.~E. Gusakov, ``{Long-lasting accretion-powered chemical heating of millisecond pulsars},'' \href{http://dx.doi.org/10.1093/mnras/stab2922}{{\em Mon. Not. Roy. Astron. Soc.} {\bfseries 508} no.~4, (2021) 6118--6127}, \href{http://arxiv.org/abs/2110.02881}{{\ttfamily arXiv:2110.02881 [astro-ph.HE]}}.

\bibitem{Bailes:1997}
M.~{Bailes}, S.~{Johnston}, J.~F. {Bell}, D.~R. {Lorimer}, B.~W. {Stappers}, R.~N. {Manchester}, A.~G. {Lyne}, L.~{Nicastro}, and B.~M. {Gaensler}, ``{Discovery of Four Isolated Millisecond Pulsars},'' \href{http://dx.doi.org/10.1086/304041}{{\em \apj} {\bfseries 481} no.~1, (May, 1997) 386--391}.

\bibitem{Reardon:2015kba}
D.~J. Reardon {\em et~al.}, ``{Timing analysis for 20 millisecond pulsars in the Parkes Pulsar Timing Array},'' \href{http://dx.doi.org/10.1093/mnras/stv2395}{{\em Mon. Not. Roy. Astron. Soc.} {\bfseries 455} no.~2, (2016) 1751--1769}, \href{http://arxiv.org/abs/1510.04434}{{\ttfamily arXiv:1510.04434 [astro-ph.HE]}}.

\bibitem{Durant:2012}
M.~{Durant}, O.~{Kargaltsev}, G.~G. {Pavlov}, P.~M. {Kowalski}, B.~{Posselt}, M.~H. {van Kerkwijk}, and D.~L. {Kaplan}, ``{The Spectrum of the Recycled PSR J0437-4715 and Its White Dwarf Companion},'' \href{http://dx.doi.org/10.1088/0004-637X/746/1/6}{{\em Astrophys. J.} {\bfseries 746} no.~1, (Feb., 2012) 6}, \href{http://arxiv.org/abs/1111.2346}{{\ttfamily arXiv:1111.2346 [astro-ph.HE]}}.

\bibitem{Gonzalez-Caniulef:2019wzi}
D.~Gonzalez-Caniulef, S.~Guillot, and A.~Reisenegger, ``{Neutron star radius measurement from the ultraviolet and soft X-ray thermal emission of PSR J0437\ensuremath{-}4715},'' \href{http://dx.doi.org/10.1093/mnras/stz2941}{{\em Mon. Not. Roy. Astron. Soc.} {\bfseries 490} no.~4, (2019) 5848--5859}, \href{http://arxiv.org/abs/1904.12114}{{\ttfamily arXiv:1904.12114 [astro-ph.HE]}}.

\bibitem{LZ:2024psa}
{\bfseries LZ} Collaboration, J.~Aalbers {\em et~al.}, ``{New constraints on ultraheavy dark matter from the LZ experiment},'' \href{http://dx.doi.org/10.1103/PhysRevD.109.112010}{{\em Phys. Rev. D} {\bfseries 109} no.~11, (2024) 112010}, \href{http://arxiv.org/abs/2402.08865}{{\ttfamily arXiv:2402.08865 [hep-ex]}}.

\end{thebibliography}\endgroup

\end{document}